\pdfoutput=1
\ifdefined\XeTeXrevision\pdfoutput=0\fi
\newif\ifarxiv
\arxivtrue

\documentclass{article}
\usepackage{submission_style,times}
\usepackage{lmodern}

\usepackage{amsmath,amsfonts,bm}

\def\eqref#1{equation~\ref{#1}}

\def\1{\bm{1}}

\DeclareMathAlphabet{\mathsfit}{\encodingdefault}{\sfdefault}{m}{sl}
\SetMathAlphabet{\mathsfit}{bold}{\encodingdefault}{\sfdefault}{bx}{n}

\usepackage{booktabs}
\usepackage{multirow}
\usepackage{array}
\usepackage{tabularx}
\usepackage{xcolor}
\usepackage{colortbl}
\usepackage{graphicx}
\graphicspath{{figures/}}
\usepackage{placeins}
\usepackage{float}
\usepackage{tikz}
\usetikzlibrary{arrows.meta,positioning,fit,calc,backgrounds}
\usepackage{pgfplots}
\definecolor{drawInk}{RGB}{35,39,47}
\definecolor{drawSlate}{RGB}{88,99,112}
\definecolor{drawGrid}{RGB}{202,208,216}
\definecolor{drawNavy}{RGB}{31,78,121}
\definecolor{drawTeal}{RGB}{38,128,128}
\definecolor{drawCoral}{RGB}{190,92,73}
\definecolor{drawAmber}{RGB}{178,128,37}
\definecolor{drawPlum}{RGB}{112,82,126}
\definecolor{plotBlue}{RGB}{0,114,178}
\definecolor{plotOrange}{RGB}{213,94,0}
\definecolor{plotGreen}{RGB}{0,158,115}
\definecolor{plotPurple}{RGB}{204,121,167}
\definecolor{plotGray}{RGB}{70,70,70}
\definecolor{drawPaper}{RGB}{250,250,248}
\definecolor{drawBlueFill}{RGB}{238,244,250}
\definecolor{drawTealFill}{RGB}{236,247,246}
\definecolor{drawWarmFill}{RGB}{250,244,236}
\definecolor{drawGrayFill}{RGB}{232,235,239}
\pgfplotsset{compat=1.17,
  drawAxis/.style={
    axis line style={draw=drawGrid},
    axis x line*=bottom,
    axis y line*=left,
    tick style={draw=drawGrid},
    tick label style={font=\scriptsize, color=drawInk},
    label style={font=\small, color=drawInk},
    grid=major,
    major grid style={draw=drawGrid!55, line width=0.25pt},
    legend style={draw=none, fill=none, font=\scriptsize},
  }
}
\newcolumntype{Y}{>{\raggedright\arraybackslash}X}
\newcolumntype{C}{>{\centering\arraybackslash}X}

\usepackage{hyperref}
\hypersetup{hidelinks}
\usepackage{url}
\providecommand{\bibfont}{}
\renewcommand{\bibfont}{\small}
\submissionfinalcopy
\author{%
\parbox{\dimexpr\textwidth-2\tabcolsep\relax}{%
\centering
\begin{tabular}{@{}c@{}}
{\normalsize \textbf{Mengzhe Geng}$^{1}$\thanks{Corresponding author.}\hspace{1.25em}\textbf{Yujia Lu}$^{2}$\hspace{1.25em}\textbf{Manuela Kunz}$^{1}$}\\[0.16em]
{\normalsize \textbf{Patrick Littell}$^{1}$}\\[0.46em]
{\normalfont\footnotesize $^{1}$National Research Council Canada \quad $^{2}$The Chinese University of Hong Kong}\\[0.10em]
{\normalfont\scriptsize \texttt{Mengzhe.Geng@nrc-cnrc.gc.ca}}
\end{tabular}%
}%
}

\title{From Scores to Evidence: Auditable Decisions Can Improve Speech Deepfake Detection}

\begin{document}
\maketitle

\fancyhead{}
\renewcommand{\headrulewidth}{0pt}
\raggedbottom
\begin{abstract}
Speech deepfake detectors usually emit one score per utterance, but a borderline score does not reveal why two examples differ during retrospective error analysis.
We ask whether a final score can be calibrated from component fields while keeping those fields visible for inspection.
We build a decision record with a passive detector score and a score from a probe applied to a marked copy.
It also includes retrieval support held out of the evaluated family, a margin from a support-set profile, and raw neighbor closeness.
A cross-fit calibrator combines these fields and two differences between raw scores into one final score.
On matched ASVspoof development data, the calibrated record reduces equal error rate (EER) by 3.48 percentage points relative to the fixed retrieval-augmented rule.
It reaches 8.43\% EER, whereas a passive WavLM baseline reaches 6.71\% on the same subset.
The record is therefore not the strongest detector in this comparison.
Its value is to retain inspectable component fields while producing one scalar score for retrospective diagnosis.
\end{abstract}

\section{Introduction}

Speech deepfakes can mimic a speaker's voice closely enough to deceive people and automated systems.
That risk has motivated extensive work on speech deepfake detection.
Yet most benchmarks and papers end with one score per utterance.
A score can rank systems, but it does not retain which component fields distinguish two borderline examples.
One example may have a large passive--retrieval raw-score gap; another may lack a probe altogether.
If only the final scalar is retained, that distinction disappears.\footnote{Code: \href{https://github.com/MENGZHEGENG/from-scores-to-evidence}{\texttt{MENGZHEGENG/from-scores-to-evidence}}; arXiv: \href{https://arxiv.org/pdf/2609.08899}{2609.08899}.}

\begin{figure*}[t]
\centering
\includegraphics[width=0.985\textwidth]{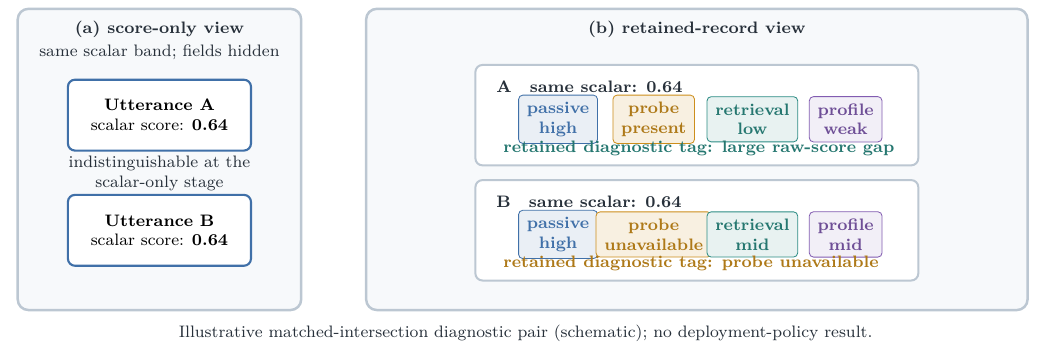}
\caption{Illustrative matched-intersection diagnostic pair. In the score-only view, two utterances share the same displayed scalar score; the retained-record view distinguishes a large raw-score gap from an unavailable probe field. The proposed decision record retains passive, probe, retrieval, and profile fields through calibration, while raw neighbor closeness remains an auxiliary diagnostic. The displayed values are schematic, and the figure does not evaluate a deployment-time inspection policy; Appendix Figure~\ref{fig:evidence-cards} shows matched-intersection examples.}
\label{fig:problem}
\end{figure*}

Figure~\ref{fig:problem} makes the diagnostic problem concrete.
Both utterances fall in the same score band, yet one has a large passive--retrieval raw-score gap and the other has an unavailable probe field.
The issue is therefore not only whether the score is high or low.
It is whether the score still retains the component fields behind it.

We retain those fields in an auditable decision record.
Here, auditable means that component fields remain visible for retrospective diagnosis; it does not validate a human audit process, source provenance, identity, or an operational inspection workflow.
Before final calibration, each utterance retains four aligned cues.
They are a passive detector score on the original audio, a keyed probe score on a marked derivative, retrieval support from a held-out reference pool, and a support-set profile margin.
We retain raw neighbor closeness as an auxiliary diagnostic.
Calibration follows assembly of the record, so the decision remains scalar while its component fields remain visible.

These streams do not mean the same thing.
The passive detector scores spoof evidence in the waveform.
The probe score is available only with a known key, a recorded embedding rule, and a derived marked input \citep{wu2025proactivepassive,ozer2026selfvoice}; otherwise, the field is unavailable and does not furnish an authentication result.
Retrieval and profile cues depend on support-pool coverage and speaker context \citep{liu2025retrievalprofile}.
The support-set profile margin does not certify identity, prompt, channel, or recording-source independence.
Early fusion treats the fields as interchangeable measurements even though they answer different retrospective diagnostic questions.

We evaluate this design on the matched ASVspoof~5 Track~1 development intersection: 4,080 utterances for which all four cues are available.
Tables~\ref{tab:matched-results} and~\ref{tab:calibration-ablation} test score performance on that intersection.
Figure~\ref{fig:evidence-card-risk} and Appendix Table~\ref{tab:record-review-load} test whether the visible cues support a retrospective, fixed-budget diagnostic; they do not evaluate a deployment policy.
WavLM is the strongest passive-only baseline on this subset.
Our question is narrower: can late calibration over aligned cues improve score performance while preserving an inspectable basis for retrospective diagnosis?

\paragraph{\textbf{Contributions.}} This paper makes three contributions.
First, it introduces an utterance-level decision record that retains aligned passive, probe, retrieval, and profile fields instead of hiding them in early fusion.
Second, a small cross-fit calibrator with two operational raw-score-gap features improves on fixed scalar fusion on the matched intersection.
Nonlinear controls leave the mechanism of this improvement unresolved.
Third, it evaluates the record through held-out-family calibration, retrospective fixed-budget diagnostics, and worked examples.
The keyed probe is an auxiliary measurement only; we make no statement about watermark robustness, attribution, or deployment-time inspection decisions.

\section{Related Work}

Table~\ref{tab:closest-work} separates the closest cited work along task and data, retained fields, evaluation target, and the boundary of the present study.
The comparison distinguishes operational decision records from artifact-specific expert semantics and from retrieval methods evaluated as stand-alone detection systems.

\begin{table}[H]
\caption{Closest-work comparison. The fields column describes values retained for a final score or diagnosis; it does not imply a human review process.}
\label{tab:closest-work}
\vspace{3pt}
\centering
\fontsize{6.05}{6.80}\selectfont
\setlength{\tabcolsep}{1.25pt}
\renewcommand{\arraystretch}{1.05}
\begin{tabularx}{\linewidth}{@{}>{\raggedright\arraybackslash}p{0.15\linewidth}>{\raggedright\arraybackslash}p{0.16\linewidth}>{\raggedright\arraybackslash}p{0.21\linewidth}>{\raggedright\arraybackslash}p{0.19\linewidth}>{\raggedright\arraybackslash}X@{}}
\toprule
{\bfseries\boldmath Work} & {\bfseries\boldmath Task / data} & {\bfseries\boldmath Method or retained fields} & {\bfseries\boldmath Evaluation target} & {\bfseries\boldmath Relationship to this study} \\
\midrule
Wu et al.\ \citep{wu2025proactivepassive} & Proactive and passive speech-deepfake detection & Passive detector and keyed-mark system scores & System-level proactive/passive comparison & Here the keyed probe is one auxiliary field; its weak stand-alone signal is reported explicitly. \\
Liu et al.\ \citep{liu2025retrievalprofile} & Zero-day detection on DeepFake-Eval-2024 & Training-free retrieval augmentation and profile matching & Detection and cross-database generalization & Here retrieval and profile values remain visible in a fixed ASVspoof~5 matched decision record. \\
Negroni et al.\ \citep{negroni2026interpretable} & Interpretable speech-deepfake detection & Trained artifact-specific experts and calibrated log-likelihood ratios & Expert semantics plus aggregate classification & Here operational raw scores and gaps remain visible; the record does not infer named artifact identities. \\
This work & 4,080 matched ASVspoof~5 development examples & Passive, keyed probe, retrieval, profile, and raw-score-gap fields & Out-of-fold score evaluation and retrospective diagnosis & Does not establish source provenance, a human review policy, or deployment utility. \\
\bottomrule
\end{tabularx}
\renewcommand{\arraystretch}{1.0}
\end{table}

\paragraph{\textbf{Speech deepfake benchmarks and detector generalization.}}
Prior automatic speaker verification (ASV) spoofing evaluations, including ASVspoof, make passive systems comparable across generator families, channels, and postprocessing conditions \citep{wang2026asvspoof5design,wang2024asvspoof5,wang2026asvspoof5evaluation,ge2025posttraining,huang2025sharpness}.
We use the official ASVspoof 5 results as benchmark context because the decision-record study relies on a smaller matched subset \citep{wang2026asvspoof5evaluation}.
RawNet2 and graph-attention spectro-temporal models remain standard passive baselines \citep{tak2021rawnet2,jung2022aasist}, while frozen self-supervised learning (SSL) systems built on wav2vec 2.0, HuBERT, and WavLM provide strong scalar references \citep{baevski2020wav2vec2,hsu2021hubert,chen2021wavlm}.
The Wav2Vec2 LV60 variant denotes the LibriVox 60k-hour pretraining configuration.
Related work on monitoring and calibration shows why ranking metrics alone are insufficient when score reliability also matters \citep{wang2026datadrift,guo2017calibration,niculescu2005probabilities}.
Our setting starts from these scalar baselines and asks what is lost when their final score hides channel-level component fields.

\paragraph{\textbf{Probe semantics and watermark context.}}
Here proactive watermarking contributes one auxiliary probe field.
Its score is meaningful only with respect to the key, embedding rule, and threat model that produced it.
Passive/proactive comparisons need metrics that preserve this semantic distinction instead of collapsing everything into a single class score \citep{wu2025proactivepassive,ge2025fakemark}.
Recent watermarking work has moved toward localized detection \citep{sanroman2024audioseal}, and self-voice conversion continues to be a demanding stress case \citep{ozer2026selfvoice}.
Shortcut learning is another concern: if the mark leaks label information, a detector may rely on the mark and ignore the spoofing trace \citep{muller2026watermarkshortcut}.
We treat the probe as an auxiliary measurement alongside the passive score, not as a replacement for it.

\paragraph{\textbf{Localization, profile matching, and interpretability.}}
Long-form and partially spoofed audio push file-level detection toward temporal forensics \citep{liu2025lensdf,zhang2024spoofdiarization,tran2026worddetection,liu2024longform}.
Retrieval and profile matching address a different limitation: a test generator, channel, or speaker relation may be absent from the supervised training set \citep{liu2025retrievalprofile,zeng2024spoofingaware}.
Work on embedding probes, spoofing-aware fusion, and expert-mixture detectors likewise motivates calibrated, interpretable score fields \citep{liu2024probingembeddings,wang2024compositionalfusion,negroni2026interpretable}.
We do not propose a new detector family.
Our contribution is a fixed record-and-review protocol in which the inputs to the late-calibrated score remain visible for retrospective diagnosis.
We frame that diagnostic stage in relation to selective classification \citep{geifman2017selective}.

\paragraph{\textbf{Positioning of the contribution.}} The cited passive, watermark, retrieval/profile, and interpretable-fusion lines motivate the individual fields, but this study does not present a new acoustic detector, a stand-alone watermark test, or a general fusion architecture.
Its unit of evaluation is a fixed record-and-diagnostic protocol.
Fields with stated semantics are retained until late calibration.
The operating score is evaluated on a matched, family-held-out intersection, and the visible fields are assessed through retrospective error ordering without a deployment-policy result.
The reported distinction is therefore the specified assembly and diagnostic evaluation, not novelty of an individual field or a semantic interpretation of a raw-score gap.

\section{Method}

Each utterance is represented by a fixed-format decision record before final calibration.
The record preserves what each stream says about the same utterance instead of replacing the streams with an early fused score.
For an input segment $x$, larger values of class-oriented fields support bona fide speech or mark consistency.
The record contains five scalar fields.
The first two are passive detector probability $s_p(x)$ on the original audio and keyed spread-spectrum probe score $s_w(x)$ on a marked derivative.
The remaining three are an inverse-distance-weighted bona fide vote $s_r(x)$ from $k=10$ support neighbors under a $k$-nearest-neighbor (kNN) rule, support-set profile-margin score $s_m(x)$ from the nearest bona fide-versus-spoof distance margin, and raw nearest-neighbor closeness $c_r(x)$.
We retain $c_r$ in its native orientation; only the cross-fit calibrator learns its orientation.

These fields have different semantics.
The passive detector scores the original utterance.
The probe is meaningful only relative to a known key, recorded embedding rule, and marked derivative; if any of these conditions is absent, the field is unavailable.
Retrieval and the support-set profile margin provide support-pool and speaker-context evidence, while $c_r$ records raw neighbor closeness without a pre-imposed class orientation.
The profile field does not identify a speaker or certify prompt, channel, or recording-source independence.
Calibration occurs only after the measurements are written side by side in the record.
Appendix Table~\ref{tab:provenance-threat-model} summarizes the resulting field-semantics and applicability boundary.

\begin{figure*}[t]
\centering
\includegraphics[width=0.985\textwidth]{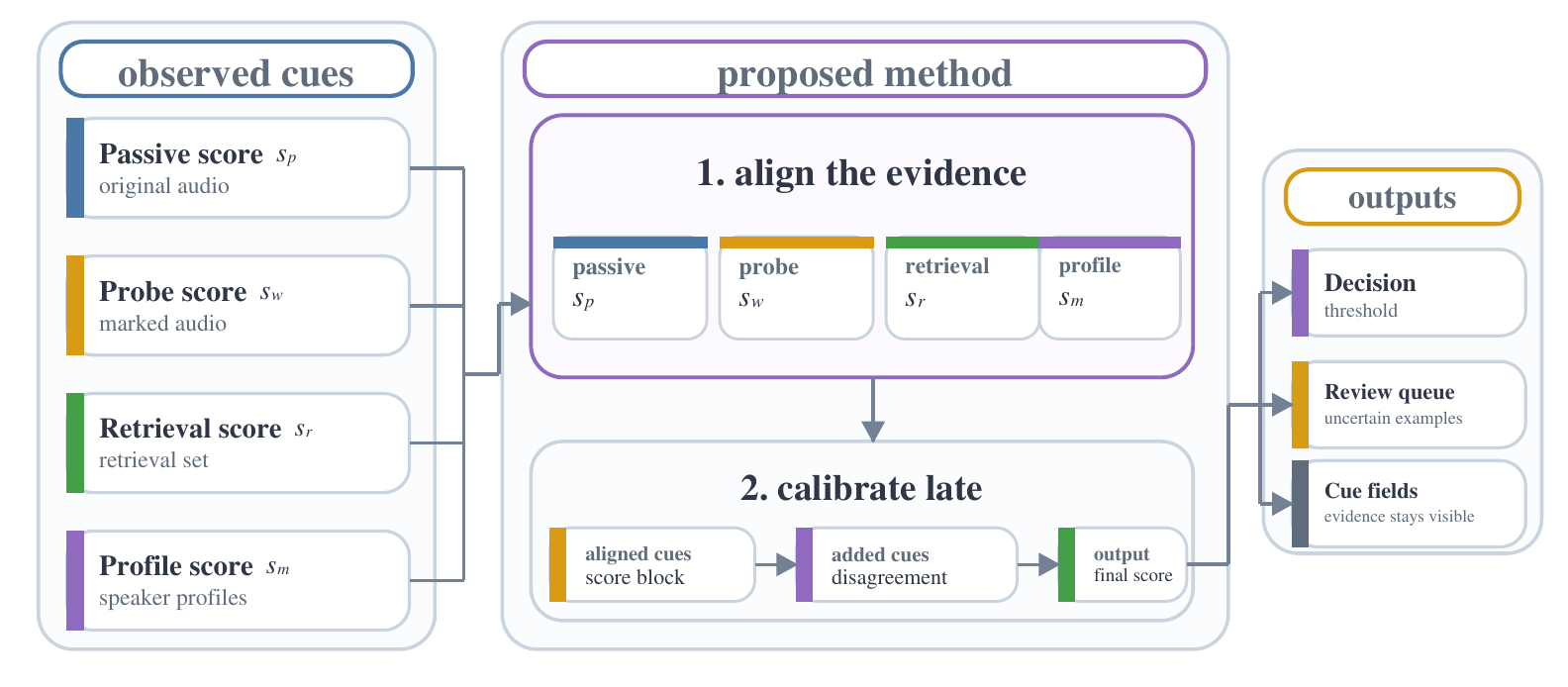}
\caption{Method overview. Four primary cue streams---passive, probe, retrieval, and profile---are first written to one utterance-level record; raw neighbor closeness is retained as an auxiliary diagnostic. The main modeling step is late calibration over a score block and two operational raw-score gap features, which yields the operating score while leaving the original cues available for thresholding and retrospective diagnostic inspection.}
\label{fig:method}
\end{figure*}

Figure~\ref{fig:method} highlights the modeling choice.
The original utterance feeds the passive detector, the marked derivative feeds the keyed probe, and the held-out support set feeds retrieval and profile scoring.
The learned step is not early fusion of all streams into one hidden score.
Instead, late calibration operates on a visible score block together with operational raw-score gap features, while thresholding and retrospective diagnostic inspection still have access to the underlying fields.

From these channels we form three fixed averages and two visible raw-score gap features.
The fixed averages are passive--watermark $f_{pw}$, retrieval-augmented fusion $f_{pwr}$, and profile-augmented fusion $f_{pwrm}$; the gap vector is $d=[|s_p-s_w|,|f_{pw}-s_r|]$.
The cross-fit calibrator is
\begin{equation}
  s_{\mathrm{rec}}(x)=\sigma\left(\beta_0+\beta_z^\top z(x)+\beta_d^\top d(x)\right),
  \label{eq:evidence-record-calibration}
\end{equation}
where $z=[s_p,s_w,f_{pw},s_r,s_m,c_r,f_{pwr},f_{pwrm}]$.
The averages are fixed before evaluation: $f_{pw}=0.5s_p+0.5s_w$, $f_{pwr}=0.5f_{pw}+0.5s_r$, and $f_{pwrm}=0.25(s_p+s_w+s_r+s_m)$.

The calibrator adds only two raw-score gap features beyond the scalar score block.
Because the averages are linear combinations of $s_p$, $s_w$, $s_r$, and $s_m$, the scalar block has rank five after $c_r$ is added.
The main ablation tests whether this operational parameterization changes performance beyond linear score fusion in the fixed feature basis.
Table~\ref{tab:calibration-ablation} reports the explicit $\beta_d=0$ ablation, passive-only transforms, no-watermark retrieval/profile controls, one-gap controls, a squared-gap parameterization, and product terms.

We use the following notation throughout.
The ``fixed retrieval-augmented rule'' is $f_{pwr}$, the ``profile-augmented rule'' is $f_{pwrm}$, and ``score fusion'' is the same cross-fit logistic protocol with $\beta_d=0$.
The learned decision-record family includes absolute-gap and squared-gap parameterizations of $s_{\mathrm{rec}}$.
Unless stated otherwise, worked examples and retrospective diagnostic analyses use the absolute-gap reporting variant because its retained coordinates are directly readable.
Its small difference from the squared-gap variant is unresolved by the paired interval.
The logistic calibrator standardizes features, initializes its intercept from the training-fold class prior, and uses an $L_2$ penalty of $10^{-3}$.
Only this out-of-fold fit learns the orientation of $c_r$; Appendix Table~\ref{tab:passive-polarity} gives an independent label-informed upper bound.
Every learned result is out of fold.
Spoof examples from a held-out synthesis family are evaluated only by a calibrator trained on the other seven families, and bona fide examples are assigned by a stable utterance-ID hash.
The final record retains component scores, the operating score, calibration bin, nearest-neighbor metadata, watermark status, and consistency flags; field retention alone does not establish source provenance, identity, or operational inspection utility.

We ask whether auxiliary streams reduce held-out-family errors and whether exposed cues annotate retrospective error ordering.
WavLM-large supplies the scalar-detector reference, and the decision record keeps complementary evidence visible.

\paragraph{\textbf{Data-handling boundary.}} The retained field list is an analysis schema, not a release specification. This manuscript does not establish permission to redistribute audio, speaker-linked identifiers, embeddings, neighbor metadata, or keys. Any reproduction artifact must follow the source-corpus terms and minimize or withhold these fields when redistribution is not authorized.

\section{Experimental Protocol}

\paragraph{\textbf{Data and splits.}}
All main experiments use ASVspoof 5 Track 1 development data.
The primary matched intersection is the utterance-identifier (ID) intersection of the passive--probe and retrieval/profile branches after holding out synthesis families A09--A16.
It contains 4,080 utterances from 705 speakers: 3,215 bona fide examples, 865 spoof examples, and 8.01 h of referenced speech.
Table~\ref{tab:protocol-details} summarizes the four fixed protocol choices: paired matched examples, a watermark-probe field with an explicit unavailability control, family-held-out retrieval support, and out-of-fold calibration.

\begin{table}[!h]
\caption{Matched held-out-family protocol used in the main comparisons.}
\label{tab:protocol-details}
\centering
\footnotesize
\setlength{\tabcolsep}{5.0pt}
\renewcommand{\arraystretch}{1.06}
\begin{tabular}{>{\raggedright\arraybackslash}p{0.19\linewidth}>{\raggedright\arraybackslash}p{0.73\linewidth}}
\toprule

{\bfseries\boldmath Component} & {\bfseries\boldmath Specification} \\
\midrule
Matched examples & ASVspoof 5 Track 1 development split; 4,080 matched examples with 4,080 unique utterances and 705 speakers (3,215 bona fide / 865 spoof); 8 held-out A09--A16 folds with 91--124 spoof examples each. Audio-path metadata covers 8.01 h over 4,080 files. \\
Passive score & Archived compact spectrogram-CNN bona fide probability for every matched comparison; the stored score is treated as a fixed input stream. No per-family score-polarity correction is applied, so family failures stay visible in Table~\ref{tab:families}. Stronger frozen-SSL passive comparisons appear in the appendix. \\
Watermark score & Spread-spectrum presence probability with one development key, strength -32.0 dB, and 16 kHz audio. The matched fusion examples all carry the same derived watermark condition across bona fide and spoof labels, so this stream tests keyed-probe behavior and a raw-score feature while keeping watermark presence independent of the class label. \\
Retrieval/profile & HuBERT-large archive; $k=10$ neighbor-label retrieval; 38,797 support embeddings and 40,000 pre-join predictions from our reimplementation, distinct from the published paper's reported scores. The identifier-overlap analysis finds 0/20,000 spoof examples with any top-10 neighbor from the held-out family, with 0 query utterance-ID matches, 0 speaker ID matches, and 0 exact audio-path matches across the top-10 list; source tags are corpus-level fields and are not used to certify prompt or recording-source separation. \\
Calibration & 8 leave-family-out folds; 10 features including passive--watermark and fusion--retrieval raw-score gaps; seed 20260821; intervals use 5,000 bootstrap resamples. \\
\bottomrule
\end{tabular}
\renewcommand{\arraystretch}{1.0}
\end{table}

The matched subset is intentionally restrictive.
We use it for paired comparisons across evidence streams and use larger score sets only to interpret coverage.
Appendix Table~\ref{tab:subset-coverage} lists example counts, speakers, families, and duration; Appendix Table~\ref{tab:retrieval-subset} reports retained proportions and retrieval EER shifts by family.
We quantify the join effect with matched-size bootstrap draws from full retrieval scores and with spoof examples omitted by the join.
No held-out family has a matched EER outside its bootstrap interval, and the largest absolute matched-minus-random EER shift is 4.30 percentage points.
Retained and omitted spoof examples differ by -0.74 on the bona fide score scale (Kolmogorov--Smirnov (KS) distance = 0.024; maximum family KS = 0.133; Appendix Table~\ref{tab:retrieval-join-selection}).
These checks bound the join effect on available retrieval scores, but they do not make the matched subset equivalent to the full retrieval distribution.
The retrieval branch therefore supplies family-held-out matched support evidence, not a source-isolated or speaker-isolated zero-day result.

Detector-strength comparisons use matched passive controls and frozen-SSL summaries kept outside the central decision-record comparison.
Outside-corpus passive analyses use public In-The-Wild, WaveFake, and ASVspoof 2021 deepfake resources \citep{muller2022generalize,liu2024longform,frank2021wavefake,liu2022asvspoof2021}.
Each family comparison reuses the same bona fide pool and swaps only the held-out spoof family.

Unless stated otherwise, EER denotes one pooled threshold over the evaluated examples.
Held-out-family EER averages evaluations that compare bona fide examples with one held-out family while reusing the full bona fide pool, and Fold EER averages the eight calibration folds.

\paragraph{\textbf{Models and scores.}}
The matched comparison uses a fixed compact convolutional neural network (CNN) on spectrograms, with four-second, 16 kHz preprocessing metadata.
This passive stream defines the main experiment.
Stronger frozen-SSL systems provide context only: Table~\ref{tab:stress-diagnostics} gives the matched WavLM-large analysis, Appendix Table~\ref{tab:attack-stress} gives frozen-SSL attack settings, and Appendix Table~\ref{tab:ssl-eval-context} gives full-evaluation passive baselines.
The watermark score is computed on a derived spread-spectrum track with one development key at $-32$ dB in the 3.0--7.6 kHz band.
Because the same derived condition is attached to bona fide and spoof labels, we interpret this score as probe behavior and as one raw-score input, not as stand-alone class evidence.
A probe field exists only for the recorded-key, recorded-embedding-rule, derived-mark condition; absent keys, absent marks, removal, corruption, and conversion are unavailable or failure conditions, not authentication outcomes.

Following the zero-day retrieval setting of \citet{liu2025retrievalprofile}, the retrieval and support-set profile channels use $k=10$ nearest neighbors after support-set feature standardization, with held-out-family support removed before scoring that family.
The main comparison fixes the HuBERT-large 20k configuration specified in Table~\ref{tab:protocol-details}: $k=10$, 38,797 support embeddings, and 40,000 pre-join predictions.
Appendix Table~\ref{tab:retrieval-sweep} lists the 13-setting sensitivity sweep, whose best setting is not used in the central matched comparison.
Appendix Table~\ref{tab:retrieval-neighbor-audit} gives the identifier-overlap audit.
The family-exclusion audit therefore reports zero held-out-family top-10 neighbors for the 20,000 spoof examples, as required by the split.
Across all 40,000 retrieval examples, it also finds no exact query, speaker, or audio-path reuse.
The available source tag is shared only at corpus granularity, so prompt, channel, and recording-source independence remain unresolved.

Retrieval choices other than the fixed main configuration are studied only as sensitivity analyses.
The sensitivity sweep does not select a result for the central matched comparison.

\paragraph{\textbf{Metrics and uncertainty.}}
We report EER, the minimum detection cost function (minDCF), and expected calibration error (ECE).
All class-oriented scores are oriented so larger values support the bona fide or mark-consistent hypothesis; the raw $c_r$ feature is handled as described above.
The main text uses a bona fide (BF)-target normalized cost ($P_{\mathrm{tar}}=0.05$, $C_{\mathrm{miss}}=1$, $C_{\mathrm{fa}}=10$), while Appendix Table~\ref{tab:cost-orientation} repeats the comparison with the target class inverted.
Because minDCF moves strongly with this prior and several settings approach the default-cost ceiling, we interpret the main comparison through EER, ECE, and paired intervals.
ECE uses 15 equal-width probability bins and is reported as an out-of-fold point estimate.
For the named calibration controls, Appendix Table~\ref{tab:calibration-ece-intervals} gives paired ECE intervals from 5,000 class-stratified resamples.
Brier values remain descriptive and are not used to rank settings.
Matched comparisons use paired class-stratified bootstrap intervals with 5,000 resamples, and learned controls are also resampled over the eight held-out folds.
Appendix Table~\ref{tab:speaker-resampled-deltas} repeats the key EER comparisons with speaker-resampled intervals.
Source-level resampling is not attempted because the matched subset exposes only one coarse source tag.

\section{Results}

\paragraph{\textbf{Main matched comparison.}}
All values in this paragraph refer to the 4,080-utterance matched intersection.
Table~\ref{tab:matched-results} fixes one HuBERT retrieval branch for the paired comparison instead of selecting the best retrieval-sweep setting.
The held-out-family EER column averages eight evaluations, each holding out one synthesis family while sharing a bona fide pool.
The WavLM entry reports the best run from a 22-run matched passive sweep; its median is 9.60\%.
The change is concentrated: A12 and A13, where the passive stream fails most often, account for most of the fixed-rule difference.
The fixed retrieval-augmented rule lowers pooled EER from 15.84\% to 11.91\% and family-macro EER from 13.85\% to 10.27\%.
Appendix Table~\ref{tab:stress-diagnostics} shows the same ordering on the WavLM-aligned subset, where EER falls from 10.36\% to 8.27\% after retrieval is added.
The fixed rule improves ranking but not calibration: ECE rises from 0.1451 to 0.2609.
The learned cross-fit record reaches 8.43\% EER with a 0.0709 ECE point estimate on the same setup.
Against the fixed retrieval-augmented rule, its paired EER difference is -3.48 percentage points (95\% CI [-4.61, -2.08]); against cross-fit scalar fusion, it is -3.48 percentage points (95\% CI [-4.60, -2.34]) (Appendix Table~\ref{tab:paired-deltas}).
This result is stronger than the WavLM sweep median, but it does not exceed the best passive WavLM run at 6.71\% EER.

The other auxiliary channels are weaker.
Nearest-distance retrieval reaches 62.68\% EER on its own, so raw closeness is not enough.
The probe is also weak as a stand-alone score: probe-only EER is 48.67\%, and passive + probe remains close to the passive CNN (21.03\% versus 21.15\%).
Appendix Table~\ref{tab:watermark-availability} shows the same pattern after removing $s_w$: pooled EER rises only slightly from 11.91\% to 12.46\%.
We retain the probe as a conditional, inspectable field.
These experiments do not establish a separable predictive benefit from the probe.

Support-set profile evidence changes the operating trade-off instead of giving a uniform gain.
Adding the profile margin lowers family-macro EER to 9.86\% and minDCF$_{\mathrm{BF}}$ to 0.7673, but pooled EER rises to 12.96\% and ECE to 0.3270.
We therefore read the matched comparison as retrieval-led improvement with a decision record that preserves raw-score gap features for later calibration and retrospective diagnostics.

\begin{table}[!t]
\caption{Matched ASVspoof 5 comparison on the 4,080-utterance intersection where passive, probe, retrieval, and profile measurements are all available on the same utterances. Lower is better. The highlighted fixed retrieval-augmented rule is the main fixed-fusion comparison; Table~\ref{tab:calibration-ablation} tests the learned operational raw-score gap features beyond these fixed combinations.}
\label{tab:matched-results}
\centering
\fontsize{6.9}{7.9}\selectfont
\setlength{\tabcolsep}{1.2pt}
\renewcommand{\arraystretch}{1.08}
\begin{tabularx}{\linewidth}{@{}>{\raggedright\arraybackslash}p{0.37\linewidth}*{4}{>{\centering\arraybackslash}X}@{}}
\toprule

{\bfseries\boldmath System} & {\bfseries\boldmath \shortstack[c]{Pooled\\EER (\%)~$\downarrow$}} & {\bfseries\boldmath \shortstack[c]{Family\\EER (\%)~$\downarrow$}} & {\bfseries\boldmath \shortstack[c]{minDCF$_{\mathrm{BF}}$~$\downarrow$}} & {\bfseries\boldmath \shortstack[c]{ECE~$\downarrow$}} \\
\midrule
Passive CNN & 21.15 & 24.44 & 1.0000 & 0.2826 \\
Passive WavLM-large (best of 22) & 6.71 & 5.33 & 0.9776 & 0.0453 \\
Watermark probe & 48.67 & 49.08 & 0.9972 & 0.4889 \\
Passive + watermark probe & 21.03 & 24.39 & 0.9984 & 0.3755 \\
Learned passive + watermark & 21.03 & 24.61 & 0.9984 & 0.0663 \\
Retrieval kNN & 15.84 & 13.85 & 0.9406 & 0.1451 \\
Distance-only retrieval & 62.68 & 65.38 & 1.0000 & 0.1708 \\
Profile margin & 34.58 & 38.01 & 0.9944 & 0.2833 \\
\rowcolor{blue!6}
\textbf{Fixed retrieval-augmented rule} & \textbf{11.91} & \textbf{10.27} & \textbf{0.8283} & \textbf{0.2609} \\
Fixed retrieval + profile & 12.96 & 9.86 & 0.7673 & 0.3270 \\
\bottomrule
\end{tabularx}
\end{table}

\paragraph{\textbf{Interpretation boundary.}} Throughout this section, ``disagreement'' and ``conflict'' name the specified raw-score gap features only; they do not put the underlying streams on a shared semantic scale or identify the source of a performance difference.
The ECE values are out-of-fold point estimates.
We use the paired intervals in Appendix Table~\ref{tab:calibration-ece-intervals} only for its named control contrasts, not to create a global calibration ranking.

Table~\ref{tab:calibration-ablation} asks what still matters once passive and retrieval scores are available.
Fold EER averages the eight calibration folds.
It should be compared with other fold-level controls in the same table, not with the held-out-family EER column in Table~\ref{tab:matched-results}.
The direct comparison is the explicit $\beta_d=0$ score-fusion ablation of Eq.~\ref{eq:evidence-record-calibration}.
This control improves on the fixed mean, reaching 11.91\% EER and a 0.0840 ECE point estimate.
The proposed record lowers ECE by 0.0130 with a paired 95\% interval of [-0.0159, -0.0090] against this named control (Appendix Table~\ref{tab:calibration-ece-intervals}).
The weaker controls identify what does not explain the gain.
Passive shape alone remains poor: $|s_p-0.5|$ gives 24.04\% EER, and $s_p$ with $s_p^2$ gives 21.50\%.
Retrieval with profile but without the passive score reaches 16.64\% EER.
Adding passive to retrieval without a watermark field reaches 11.67\% EER, indicating that most class information already lies in the passive and retrieval channels.
The remaining difference tracks the raw-feature parameterization, not a strong independent probe score.
The no-watermark scalar expansion with $s_p^2$ and $|s_p-0.5|$ reaches 8.67\% EER; its paired difference from the full model is -0.24 percentage points [-0.81, +0.46].
The two single-gap controls reach 8.90\% for $|s_p-s_w|$ and 8.78\% for $|f_{pw}-s_r|$.
As a negative control, fold shuffling breaks the example-level link in $s_w$ while preserving each fold's watermark-score distribution.
It reaches 8.78\% EER, only 0.35 percentage points from the full model.
This result points again to score geometry or direct class evidence in the watermark field.

The learned record uses an operational raw-score-gap parameterization.
The squared-gap model gives the lowest point estimate at 8.33\% EER and 0.0709 ECE.
We use the absolute-gap reporting variant in the main analysis because its coordinates are easier to read in worked examples and retrospective diagnostics.
Its 0.10 percentage-point EER gap to the squared variant is not resolved by the paired interval in Appendix Table~\ref{tab:paired-deltas}.
Its paired ECE difference is also unresolved at +0.0001 [-0.0016, +0.0021] (Appendix Table~\ref{tab:calibration-ece-intervals}).
The no-watermark nonlinear control is also near the full model (8.67\% EER; paired difference -0.24 percentage points [-0.81, +0.46]).
Speaker-resampled intervals in Appendix Table~\ref{tab:speaker-resampled-deltas} favor the absolute-gap model over the product-term nonlinear control (-1.72 percentage points [-2.72, -0.65]), while leaving near-tie settings unresolved.
These controls do not isolate raw-score gaps as the source of the performance change.

\begin{table}[!t]
\caption{Cross-fit calibration ablation on the 4,080 matched examples. Pooled EER uses one global threshold. The controls test the specified operational raw-score gap features against generic reshaping of passive and retrieval scores; they do not identify the source of a performance difference.}
\label{tab:calibration-ablation}
\centering
\fontsize{6.65}{7.45}\selectfont
\setlength{\tabcolsep}{0.8pt}
\renewcommand{\arraystretch}{1.05}
\begin{tabularx}{0.96\linewidth}{@{}>{\raggedright\arraybackslash}p{0.22\linewidth}>{\raggedright\arraybackslash}X*{5}{>{\centering\arraybackslash}X}@{}}
\toprule
{\bfseries\boldmath Setting} & {\bfseries\boldmath Feature set} & {\bfseries\boldmath \shortstack[c]{EER\\(\%)~$\downarrow$}} & {\bfseries\boldmath \shortstack[c]{Fold EER\\(\%)~$\downarrow$}} & {\bfseries\boldmath \shortstack[c]{minDCF$_{\mathrm{BF}}$\\$\downarrow$}} & {\bfseries\boldmath ECE~$\downarrow$} & {\bfseries\boldmath Brier~$\downarrow$} \\
\midrule
Fixed retrieval rule & fixed mean & 11.91 & 10.04 & 0.8283 & 0.2609 & 0.1415 \\
Cross-fit scalar fusion & $\beta_d=0$ scalar scores & 11.91 & 9.23 & 0.9107 & 0.0840 & 0.0797 \\
Passive margin only & $|s_p-0.5|$ & 24.04 & 23.92 & 0.9991 & 0.2064 & 0.1824 \\
Passive shape control & $s_p$ plus $s_p^2$ & 21.50 & 24.14 & 1.0000 & 0.2305 & 0.1451 \\
Retrieval + profile only & no probe & 16.64 & 13.58 & 0.9978 & 0.1385 & 0.1129 \\
Passive + retrieval only & no probe & 11.67 & 9.26 & 0.9238 & 0.0904 & 0.0797 \\
No-probe nonlinear control & passive + retrieval shape & 8.67 & 7.34 & 0.9984 & 0.0756 & 0.0645 \\
Passive-probe gap & scalar + $|s_p-s_w|$ & 8.90 & 7.99 & 0.9611 & 0.0761 & 0.0680 \\
Shuffled-probe control & fold-shuffled $W$ + gap & 8.78 & 7.99 & 0.9652 & 0.0759 & 0.0673 \\
Fusion-retrieval gap & scalar + $|f_{pw}-s_r|$ & 8.78 & 7.30 & 0.8591 & 0.0752 & 0.0691 \\
Squared-gap record & squared gaps & 8.33 & 6.78 & 0.9742 & 0.0709 & 0.0631 \\
Product-term control & score products & 10.16 & 8.00 & 0.8986 & 0.0780 & 0.0751 \\
\rowcolor{blue!6}
\textbf{Proposed decision record} & \textbf{absolute gaps} & \textbf{8.43} & \textbf{7.21} & \textbf{0.9365} & \textbf{0.0709} & \textbf{0.0640} \\
\bottomrule
\end{tabularx}
\renewcommand{\arraystretch}{1.0}
\end{table}

The absolute-gap form is retained for the worked records and retrospective diagnostics because its coordinates are directly readable.
This reporting convention does not make it a confirmed score winner: the squared-gap and no-watermark nonlinear variants remain nearby, and the reported controls do not isolate a semantic interaction among streams.

\paragraph{\textbf{Where performance changes.}} The held-out-family results show that the improvement is concentrated, not uniform.
Relative to cross-fit score fusion, the absolute-gap model lowers mean Fold EER by 2.02 percentage points, from 9.23\% to 7.21\%.
Per-family changes range from -0.12 percentage points on A14 to +6.57 percentage points on A12, and A11--A12 account for 1.46 percentage points of the mean reduction.

A12 illustrates where the absolute-gap parameterization differs most from scalar fusion.
Appendix Table~\ref{tab:passive-polarity} shows that spoof examples receive higher passive bona fide probabilities than bona fide examples on average (0.945 versus 0.627).
The passive score consequently yields 81.31\% EER, whereas retrieval is correctly oriented at 2.03\%.
In Table~\ref{tab:calibration-folds}, the absolute-gap model lowers the same family's Fold EER from 13.01\% to 6.44\%.
The split-stability check supports this association.
Appendix Table~\ref{tab:calibration-split-stability} repeats learned calibrators over ten alternate bona fide assignments while keeping spoof folds fixed.
The absolute-gap model stays within 8.33\%--8.55\% pooled EER, compared with 11.69\%--12.05\% for score fusion.
These results associate the largest performance change with a passive failure mode; they do not identify its mechanism.

\paragraph{\textbf{Passive-reference context.}} The stronger passive baseline narrows the interpretation.
Under the same cross-fit protocol on the same WavLM-aligned examples, the WavLM decision-record model reaches 9.14\% EER, compared with 11.79\% for passive calibration alone.
Against the broader passive sweep, however, the best of 22 WavLM-large runs still reaches 6.71\% EER, while the sweep median is 9.60\%.
The record therefore improves on the same-protocol passive control and stays below the sweep median, but it does not exceed the best passive run.
At fold level, the ordering reverses: the record reaches 6.56\% EER versus 5.00\% for passive cross-fit calibration.
We use WavLM as the scalar reference for exactly that reason.

On the WavLM-aligned subset, passive scoring yields 22.79\% EER, retrieval yields 10.36\%, and the record plus retrieval yields 8.27\%.
Appendix Tables~\ref{tab:watermark-operating-counts} and~\ref{tab:stress-diagnostics} support the same narrow interpretation of the probe: direct marked/unmarked pairs separate cleanly, whereas source-pair and self-voice-conversion controls remain near chance.
The probe is therefore a conditional measurement tied to the marked derivative, key, and threat model, not a stand-alone detector.

\paragraph{\textbf{Retrospective diagnostic behavior.}} The analysis is retrospective: each system's EER threshold is estimated from the evaluated matched set, as stated in Table~\ref{tab:evidence-card-triage}.
It asks which score places observed mistakes nearest the top of a fixed review queue, not whether a deployment-time inspection policy succeeds.
In the HuBERT retrieval-and-profile setting, the cross-fit absolute-gap record ranks its observed errors most effectively.
In the first 10\% of reviewed examples, it captures 47.97\% of its own errors, compared with 38.89\% for the fixed retrieval-augmented rule and 29.26\% for retrieval kNN (Table~\ref{tab:evidence-card-triage}).

This is a ranking result, not a restatement of pooled EER.
After deferring the first 10\%, the residual error rate falls to 4.87\%, compared with 8.09\% for the fixed retrieval-augmented rule and 12.45\% for retrieval kNN.
The fixed rule is slightly more precise at the same budget (46.32\% versus 40.44\%), but the learned record pulls more total mistakes toward the inspection boundary.

The WavLM analysis draws the same distinction between detection and retrospective error ordering.
The best passive WavLM run reaches 6.72\% full-set error.
Adding auxiliary fields does not improve that decision rule, but the corresponding decision-record model still captures 50.40\% of its own errors in the first 10\% of reviewed examples.
The record therefore orders observed errors differently even when it is not the best detector.

\begin{figure}[t]
\centering
\begin{minipage}[t]{0.97\linewidth}
\centering
\begin{minipage}[t]{0.42\linewidth}
\centering
\makebox[\linewidth][c]{%
\begin{tikzpicture}
\begin{axis}[
  drawAxis,
  width=0.92\linewidth,
  height=4.45cm,
  scale only axis,
  xmode=log,
  xmin=0.018,
  xmax=1.10,
  ymin=0,
  ymax=1.04,
  title={(a) Retrospective diagnostic},
  title style={at={(0.5,1.14)}, anchor=south, font=\small, color=drawInk},
  xlabel={inspection load (\%, log scale)},
  ylabel={errors captured (\%)},
  xtick={0.02,0.05,0.1,0.2,0.5,1.0},
  xticklabels={2,5,10,20,50,100},
  ytick={0,0.2,0.5,0.8,1.0},
  yticklabels={0,20,50,80,100},
  xmajorgrids=true,
  ymajorgrids=true,
  clip=false,
  tick align=outside,
  xlabel style={font=\small, yshift=0.2ex},
  ylabel style={font=\small, yshift=-0.4ex},
  tick label style={font=\scriptsize},
  legend style={
    at={(0.5,1.01)},
    anchor=south,
    legend columns=2,
    draw=drawGrid!68,
    fill=white,
    fill opacity=0.97,
    text opacity=1,
    rounded corners=1pt,
    font=\scriptsize,
    inner xsep=2.0pt,
    inner ysep=0.9pt,
    cells={anchor=west},
    /tikz/every even column/.append style={column sep=0.45em},
  },
  legend image post style={line width=0.60pt,mark size=0.82pt},
  every axis plot/.append style={line join=round, line cap=round},
]
\addplot+[line width=0.72pt, solid, mark=*, mark size=0.88pt, draw=plotBlue!88!black, mark options={fill=white, draw=plotBlue!88!black, line width=0.18pt}] table[x=coverage,y=low_margin,col sep=comma] {figures/evidence_card_risk_nonzero.dat};
\addlegendentry{margin}
\addplot+[line width=0.68pt, dash pattern=on 2.1pt off 2.1pt, mark=square*, mark size=0.84pt, draw=plotOrange!88!black, mark options={fill=white, draw=plotOrange!88!black, line width=0.18pt}] table[x=coverage,y=high_disagreement,col sep=comma] {figures/evidence_card_risk_nonzero.dat};
\addlegendentry{raw-score gap}
\addplot+[line width=0.68pt, dash pattern=on 0.95pt off 1.25pt on 0.22pt off 1.25pt, mark=triangle*, mark size=0.86pt, draw=plotGreen!82!black, mark options={fill=white, draw=plotGreen!82!black, line width=0.18pt}] table[x=coverage,y=high_distance,col sep=comma] {figures/evidence_card_risk_nonzero.dat};
\addlegendentry{distance}
\addplot+[line width=0.56pt, densely dotted, mark=diamond*, mark size=0.78pt, draw=plotGray!72, mark options={fill=white, draw=plotGray!72, line width=0.16pt}] table[x=coverage,y=random,col sep=comma] {figures/evidence_card_risk_nonzero.dat};
\addlegendentry{random}
\end{axis}
\end{tikzpicture}%
}
\end{minipage}\hspace{0.13\linewidth}%
\begin{minipage}[t]{0.42\linewidth}
\centering
\makebox[\linewidth][c]{%
\begin{tikzpicture}
\begin{axis}[
  drawAxis,
  width=0.92\linewidth,
  height=4.45cm,
  scale only axis,
  xmin=0,
  xmax=1.02,
  ymin=0,
  ymax=1.04,
  title={(b) Calibration},
  title style={at={(0.5,1.14)}, anchor=south, font=\small, color=drawInk},
  xlabel={mean score (\%)},
  ylabel={empirical bona fide rate (\%)},
  xtick={0,0.2,0.5,0.8,1.0},
  xticklabels={0,20,50,80,100},
  ytick={0,0.2,0.5,0.8,1.0},
  yticklabels={0,20,50,80,100},
  xmajorgrids=true,
  ymajorgrids=true,
  clip=false,
  tick align=outside,
  xlabel style={font=\small, yshift=0.2ex},
  ylabel style={font=\small, yshift=-0.4ex},
  tick label style={font=\scriptsize},
  legend style={
    at={(0.5,1.03)},
    anchor=south,
    legend columns=2,
    draw=drawGrid!68,
    fill=white,
    fill opacity=0.97,
    text opacity=1,
    rounded corners=1pt,
    font=\scriptsize,
    inner xsep=2.0pt,
    inner ysep=0.9pt,
    cells={anchor=west},
    /tikz/every even column/.append style={column sep=0.34em},
  },
  legend image post style={line width=0.60pt,mark size=0.82pt},
  every axis plot/.append style={line join=round, line cap=round},
]
\addplot+[line width=0.58pt, dash pattern=on 2.0pt off 1.8pt, mark=none, draw=plotGray!64] coordinates {(0,0) (1,1)};
\addlegendentry{ideal}
\addplot+[line width=0.68pt, dash pattern=on 2.1pt off 2.1pt, mark=square*, mark size=0.84pt, draw=plotOrange!88!black, mark options={fill=white, draw=plotOrange!88!black, line width=0.18pt}] table[x=mean_score,y=empirical,col sep=comma] {figures/reliability_retrieval.dat};
\addlegendentry{retrieval kNN}
\addplot+[line width=0.72pt, solid, mark=*, mark size=0.88pt, draw=plotBlue!88!black, mark options={fill=white, draw=plotBlue!88!black, line width=0.18pt}] table[x=mean_score,y=empirical,col sep=comma] {figures/reliability_crossfit.dat};
\addlegendentry{cross-fit model}
\end{axis}
\end{tikzpicture}%
}
\end{minipage}%

\vspace{5.0ex}

\begin{minipage}[t]{0.42\linewidth}
\centering
\makebox[\linewidth][c]{%
\begin{tikzpicture}
\begin{axis}[
  drawAxis,
  width=0.92\linewidth,
  height=3.25cm,
  scale only axis,
  xbar,
  bar width=4.6pt,
  xmin=0,
  xmax=0.95,
  symbolic y coords={Passive,Retrieval,Fixed-rule,Learned-record,Cue union},
  ytick=data,
  y dir=reverse,
  enlarge y limits=0.18,
  title={(c) Retrospective same-load},
  title style={font=\small, color=drawInk, yshift=-0.4ex},
  xlabel={percentage (\%)},
  xtick={0,0.2,0.4,0.6,0.8},
  xticklabels={0,20,40,60,80},
  xmajorgrids=true,
  ymajorgrids=false,
  clip=false,
  tick align=outside,
  xlabel style={font=\small, yshift=0.2ex},
  tick label style={font=\scriptsize},
  yticklabel style={font=\scriptsize},
  legend style={
    at={(0.5,-0.34)},
    anchor=north,
    legend columns=2,
    draw=drawGrid!68,
    fill=white,
    fill opacity=0.97,
    text opacity=1,
    rounded corners=1pt,
    font=\scriptsize,
    inner xsep=2.0pt,
    inner ysep=0.9pt,
    cells={anchor=west},
    /tikz/every even column/.append style={column sep=0.42em},
  },
  legend image post style={line width=0.60pt,mark size=0.82pt},
]
\addplot+[xbar, fill=plotBlue!24, draw=plotBlue!86!black, line width=0.36pt] table[x=coverage,y=queue_short,col sep=comma] {figures/review_record_same_load.dat};
\addlegendentry{error coverage}
\addplot+[only marks, mark=diamond*, mark size=1.42pt, draw=plotOrange!88!black, mark options={fill=white, draw=plotOrange!88!black, line width=0.22pt}] table[x=precision,y=queue_short,col sep=comma] {figures/review_record_same_load.dat};
\addlegendentry{precision}
\end{axis}
\end{tikzpicture}%
}
\end{minipage}\hspace{0.13\linewidth}%
\begin{minipage}[t]{0.42\linewidth}
\centering
\makebox[\linewidth][c]{%
\begin{tikzpicture}
\begin{axis}[
  drawAxis,
  width=0.92\linewidth,
  height=3.25cm,
  scale only axis,
  xbar,
  bar width=4.6pt,
  xmin=0,
  xmax=0.70,
  symbolic y coords={Multi-cue,Passive only,Retrieval only,Threshold only},
  ytick=data,
  y dir=reverse,
  enlarge y limits=0.18,
  title={(d) Cue overlap},
  title style={font=\small, color=drawInk, yshift=-0.4ex},
  xlabel={covered errors (\%)},
  xtick={0,0.2,0.4,0.6},
  xticklabels={0,20,40,60},
  xmajorgrids=true,
  ymajorgrids=false,
  clip=false,
  tick align=outside,
  xlabel style={font=\small, yshift=0.2ex},
  tick label style={font=\scriptsize},
  yticklabel style={font=\scriptsize},
]
\addplot+[xbar, fill=plotGreen!24, draw=plotGreen!82!black, line width=0.36pt] table[x=share,y=reason,col sep=comma] {figures/review_record_overlap.dat};
\end{axis}
\end{tikzpicture}%
}
\end{minipage}
\end{minipage}
\vspace{4.2ex}
\caption{Retrospective matched-example calibration and diagnostic analysis. (a) Each curve gives the share of fixed-rule threshold errors recovered within a retrospective inspection load; post-deferral error and AURC are summarized in Appendix Table~\ref{tab:evidence-card-triage}. (b) Fifteen equal-width reliability bins compare retrieval kNN with the cross-fit model. (c) Every queue flags 33.75\% of matched examples; the four-cue union retains most of the cross-fit margin queue's error coverage with a small precision drop while preserving explicit diagnostic cues. (d) Most covered errors trigger more than one cue, and the single-cue remainder is dominated by passive-record decision mismatch. Exact fixed-budget values are listed in Appendix Table~\ref{tab:record-review-load}. This figure does not evaluate a deployment-time inspection policy.}
\label{fig:evidence-card-risk}
\end{figure}
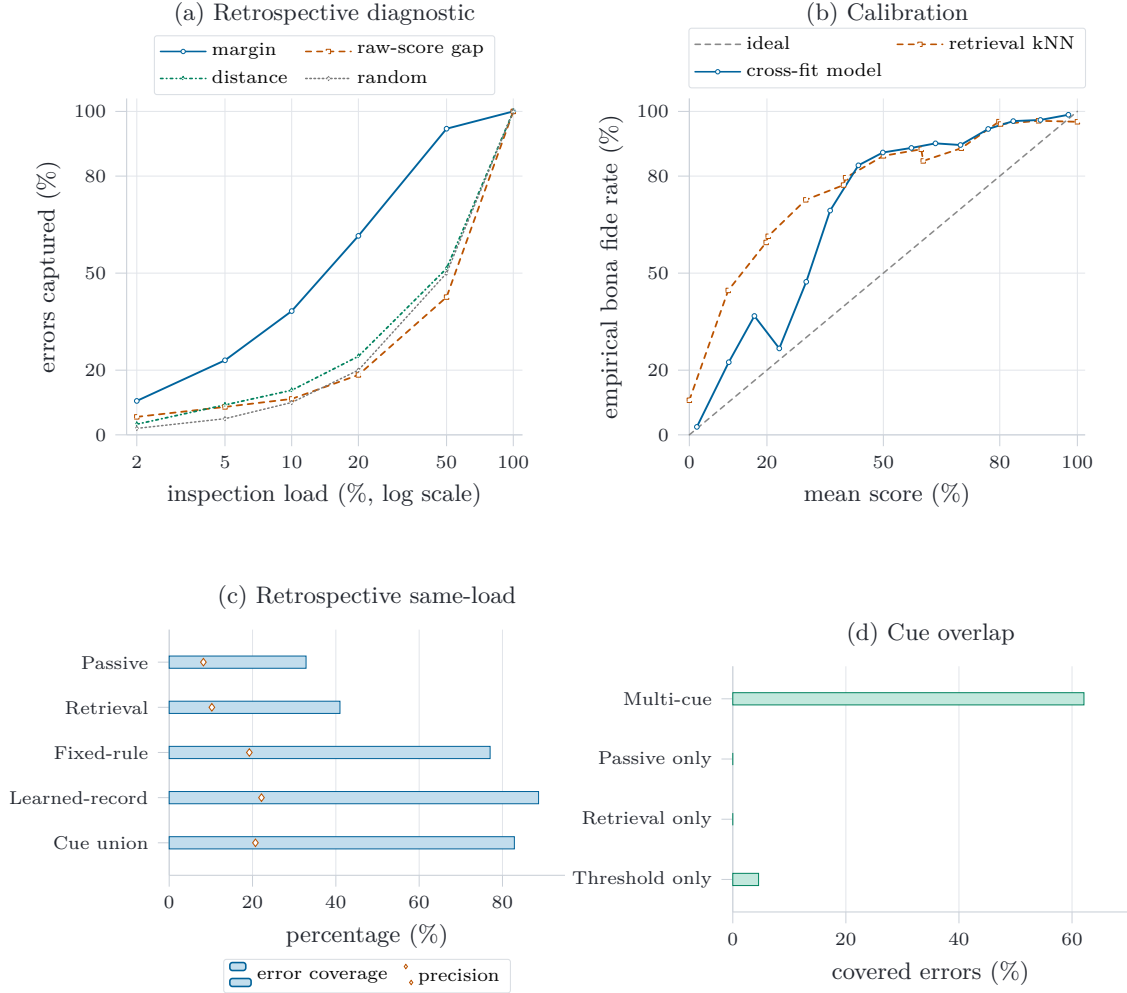

Figure~\ref{fig:evidence-card-risk} separates three retrospective diagnostic questions: which cues help before calibration, how well the learned score is calibrated, and whether visible cues still help at a fixed inspection load.
In panel (a), threshold margin is the strongest pre-calibration triage signal.
At a 20\% retrospective inspection load, it captures 61.5\% of fixed-rule errors.
Nearest-neighbor distance captures 24.3\%, and the raw-score gap captures 18.5\%, compared with 20.0\% for a random queue.
Margin therefore gives the clearest pre-calibration ordering; distance and the raw-score gap remain secondary warning signals.

Panel (b) uses the same 15 equal-width bins as the ECE calculation in Tables~\ref{tab:matched-results} and~\ref{tab:calibration-ablation}.
All 15 cross-fit bins sit above the diagonal.
This indicates mild under-confidence on bona fide examples; the pattern is not a severe shape failure.
The largest gap occurs in the bin centered near 0.436, where the mean score is 0.436 and the empirical bona fide rate is 0.833.
Appendix Table~\ref{tab:calibration-reliability} repeats the calculation with equal-mass bins; for the cross-fit model, ECE changes only slightly, from 0.0709 to 0.0715.
A foldwise isotonic refit lowers it further to 0.0178.

The fixed-budget comparison asks whether record-derived cues retain inspectable reasons for retrospective inspection after the score has been learned.
Appendix Table~\ref{tab:evidence-card-record-diagnostic} defines these four cues: record-threshold proximity, passive--record decision mismatch, retrieval--record decision mismatch, and a record--retrieval gap.
Figure~\ref{fig:evidence-card-risk}(c) compares them at a fixed budget, Figure~\ref{fig:evidence-card-risk}(d) breaks down covered errors, and Appendix Table~\ref{tab:record-review-load} gives the exact values.
At the shared 1,377-example retrospective inspection load (33.75\%), their union covers 82.85\% of the calibrated rule's observed errors.
A passive-margin queue covers 32.85\%, and a retrieval-margin queue covers 40.99\%.
The learned-record margin queue is stronger on both error coverage (88.66\%) and precision (22.15\%).
The four-cue union trades those quantities for multiple explicit diagnostic reasons; it covers 82.85\% of errors with 20.70\% precision.
This analysis does not establish that the cue union improves a deployment-time review policy.

\begin{figure*}[t]
\centering
\begin{tikzpicture}[
  font=\scriptsize\sffamily,
  casecard/.style={line width=0.68pt, rounded corners=5pt, minimum width=6.62cm, minimum height=3.50cm, inner sep=0pt, align=left},
  cardtitle/.style={font=\scriptsize\sffamily\bfseries, text=drawInk},
  statuslabel/.style={font=\scriptsize\sffamily\bfseries},
  cardtext/.style={font=\scriptsize\sffamily, text=drawInk},
  cardnote/.style={font=\scriptsize\sffamily\bfseries},
  barlabel/.style={font=\scriptsize\sffamily, text=drawInk},
  barvalue/.style={font=\scriptsize\sffamily, text=drawSlate},
  barframe/.style={draw=drawGrid!76, fill=white, line width=0.25pt}
]

\begin{scope}[shift={(-3.45,0.00)}]
\node[casecard, fill=drawTealFill, draw=drawTeal!82!black] (card) at (0,0) {};
\node[cardtitle, anchor=north west] at (-3.04,1.46) {\textbf{C1. retrieval-rescued spoof}};
\node[statuslabel, anchor=north east, text=drawTeal!82!black] at (3.04,1.46) {\textbf{correct}};
\node[cardtext, anchor=north west] at (-3.04,1.10) {truth: spoof $\rightarrow$ spoof};
\node[cardtext, anchor=north west] at (-3.04,0.82) {support: fold A12; mark present};
\node[cardtext, anchor=north west] at (-3.04,0.56) {nearest: NN A08 ($d=19.2$)};

\node[barlabel, anchor=east] at (-2.70,0.10) {$s_p$};
\draw[barframe] (-2.62,0.06) rectangle ++(1.36,0.09);
\fill[drawNavy!82] (-2.62,0.06) rectangle ++(1.36,0.09);
\node[barvalue, anchor=west] at (-1.18,0.10) {1.00};

\node[barlabel, anchor=east] at (-2.70,-0.16) {$s_w$};
\draw[barframe] (-2.62,-0.21) rectangle ++(1.36,0.09);
\fill[drawAmber!88!black] (-2.62,-0.21) rectangle ++(0.57,0.09);
\node[barvalue, anchor=west] at (-1.18,-0.16) {0.42};

\node[barlabel, anchor=east] at (0.36,0.10) {$s_r$};
\draw[barframe] (0.44,0.06) rectangle ++(1.36,0.09);
\fill[drawTeal!88!black] (0.44,0.06) rectangle ++(0.00,0.09);
\node[barvalue, anchor=west] at (1.88,0.10) {0.00};

\node[barlabel, anchor=east] at (0.36,-0.16) {$s_m$};
\draw[barframe] (0.44,-0.21) rectangle ++(1.36,0.09);
\fill[drawPlum!78!black] (0.44,-0.21) rectangle ++(0.68,0.09);
\node[barvalue, anchor=west] at (1.88,-0.16) {0.50};
\node[cardtext, anchor=north west] at (-3.04,-0.58) {fixed 0.35; calibrated 0.28; bin 5/15};
\node[cardtext, anchor=north west] at (-3.04,-0.88) {gaps: $|s_p-s_w|$ 0.58; $|f_{pw}-s_r|$ 0.71};
\node[cardnote, anchor=north west, text=drawTeal!82!black] at (-3.04,-1.22) {cue: retrieval rescue};
\end{scope}

\begin{scope}[shift={(3.45,0.00)}]
\node[casecard, fill=drawWarmFill, draw=drawAmber!88!black] (card) at (0,0) {};
\node[cardtitle, anchor=north west] at (-3.04,1.46) {\textbf{C2. retrieval failure}};
\node[statuslabel, anchor=north east, text=drawAmber!88!black] at (3.04,1.46) {\textbf{error}};
\node[cardtext, anchor=north west] at (-3.04,1.10) {truth: spoof $\rightarrow$ bona fide};
\node[cardtext, anchor=north west] at (-3.04,0.82) {support: fold A13; mark present};
\node[cardtext, anchor=north west] at (-3.04,0.56) {nearest: NN bona fide ($d=30.0$)};

\node[barlabel, anchor=east] at (-2.70,0.10) {$s_p$};
\draw[barframe] (-2.62,0.06) rectangle ++(1.36,0.09);
\fill[drawNavy!82] (-2.62,0.06) rectangle ++(0.09,0.09);
\node[barvalue, anchor=west] at (-1.18,0.10) {0.07};

\node[barlabel, anchor=east] at (-2.70,-0.16) {$s_w$};
\draw[barframe] (-2.62,-0.21) rectangle ++(1.36,0.09);
\fill[drawAmber!88!black] (-2.62,-0.21) rectangle ++(0.40,0.09);
\node[barvalue, anchor=west] at (-1.18,-0.16) {0.30};

\node[barlabel, anchor=east] at (0.36,0.10) {$s_r$};
\draw[barframe] (0.44,0.06) rectangle ++(1.36,0.09);
\fill[drawTeal!88!black] (0.44,0.06) rectangle ++(1.36,0.09);
\node[barvalue, anchor=west] at (1.88,0.10) {1.00};

\node[barlabel, anchor=east] at (0.36,-0.16) {$s_m$};
\draw[barframe] (0.44,-0.21) rectangle ++(1.36,0.09);
\fill[drawPlum!78!black] (0.44,-0.21) rectangle ++(0.68,0.09);
\node[barvalue, anchor=west] at (1.88,-0.16) {0.50};
\node[cardtext, anchor=north west] at (-3.04,-0.58) {fixed 0.59; calibrated 0.64; bin 10/15};
\node[cardtext, anchor=north west] at (-3.04,-0.88) {gaps: $|s_p-s_w|$ 0.23; $|f_{pw}-s_r|$ 0.82};
\node[cardnote, anchor=north west, text=drawAmber!88!black] at (-3.04,-1.22) {cue: large raw-score gap};
\end{scope}
\end{tikzpicture}%
\caption{Worked matched-intersection examples from the main decision-record model. C1 is a correct spoof from family A12: the passive stream is near the bona fide side, but retrieval support drives the calibrated score below threshold. C2 is a failure from family A13: passive evidence points toward spoof, retrieval points strongly toward bona fide, and a large raw-score gap remains visible in the final record.}
\label{fig:worked-cases}
\end{figure*}
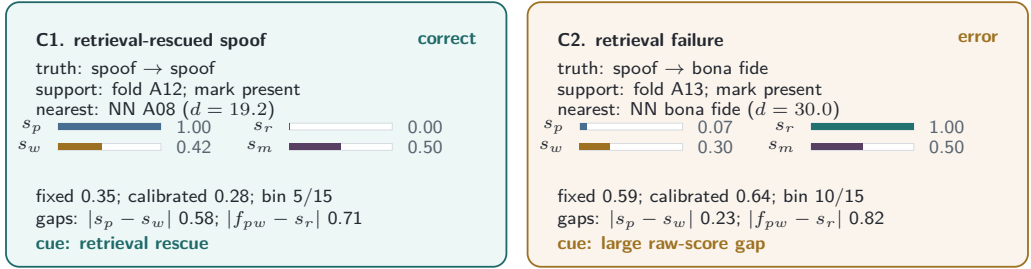

Among the 285 covered errors, 177 trigger at least two cues (62.1\%), as shown in Figure~\ref{fig:evidence-card-risk}(d).
Single-cue cases are mostly ordinary near-threshold decisions: 13 are flagged only by threshold proximity.
Figure~\ref{fig:worked-cases} shows two matched examples that make the record concrete: one retrieval-rescued spoof and one retrieval-driven failure.
In the worked records, NN denotes the nearest neighbor.

Figure~\ref{fig:worked-cases} gives two matched examples.
C1 is a correctly identified spoof from family A12.
Its passive score remains on the bona fide side (1.00), but retrieval support falls to 0.00 and moves the calibrated score below threshold, to 0.28.
C2 shows the opposite failure in family A13: passive evidence points toward spoof (0.07), retrieval points toward bona fide (1.00), and the raw-score gap remains large ($|f_{pw}-s_r|=0.82$) although the final decision is wrong.

The paired bootstrap intervals preserve the same ordering.
The fixed retrieval-augmented rule reduces pooled EER by 9.24 percentage points relative to the passive CNN and by 3.92 percentage points relative to retrieval-only evidence (Appendix Table~\ref{tab:paired-deltas}).
The learned absolute-gap model has lower EER than score fusion, the one-conflict controls, and the product-term nonlinear control, while remaining statistically tied with the squared-gap variant.
Appendix Table~\ref{tab:cost-orientation} shows that minDCF$_{\mathrm{BF}}$ still favors the fixed retrieval-augmented rule (0.8283 versus 0.9365).
We therefore use the fixed rule for threshold-oriented discussion and use learned models only for retrospective ranking, calibration, and diagnostic analyses.

\paragraph{\textbf{Family-level variation.}}
Family-level comparisons show that the pooled gain is concentrated, not uniform.
Against the passive CNN, the fixed retrieval-augmented rule improves on A12, A13, A14, and A16, is unresolved on A09, A10, and A15, and regresses on A11.
A12 is the largest single-family contributor in this error accounting.
Appendix Table~\ref{tab:passive-polarity} shows that a label-informed polarity correction lowers family-macro passive EER from 24.44\% to 15.86\%.

The error accounting tells the same story.
Bona fide examples contribute 297 of the 377 net fewer errors, A12 and A13 contribute 90 and 38, and A11 plus A15 offset the gain by 30 and 12 errors.
If A12 and A13 are removed, the family-macro EERs become nearly identical (10.20\% for the passive CNN and 10.28\% for the fixed retrieval-augmented rule).
We therefore interpret the pooled improvement as recovery from a small set of passive failure modes, not as a uniform advantage over the passive stream.
Appendix Figure~\ref{fig:family-eer} and Appendix Tables~\ref{tab:passive-card-error-accounting} and~\ref{tab:family-influence} give the full breakdown and paired intervals.

\FloatBarrier

\paragraph{\textbf{Component ablations.}}
The ablations indicate that retrieval helps through neighbor labels, not through raw distance alone.
The fixed retrieval-augmented rule lowers EER by 3.92 percentage points relative to retrieval kNN in the fixed HuBERT setting, whereas nearest-distance retrieval reaches 62.68\% EER.
Out-of-fold calibration recovers some information from this inverted distance signal, but the fixed rule does not.

The keyed probe is a conditional measurement, not a stand-alone class score: direct marked/unmarked pairs are separable, whereas matched and conversion controls carry little class information in the probe score itself.
In the WavLM setting, the passive detector is the accuracy reference.
The decision record exposes the auxiliary raw-score fields side by side, but these experiments do not isolate a predictive contribution from the probe.

The squared-gap model gives the best point estimate in Table~\ref{tab:calibration-ablation}.
It reaches 8.33\% EER and 0.0709 ECE, but Appendix Table~\ref{tab:paired-deltas} does not resolve its 0.10 percentage-point EER difference from the absolute-gap model.
We therefore keep the absolute-gap reporting variant in the main analysis because its raw-score coordinates are easier to inspect in worked examples and retrospective diagnostic settings.
The one-gap, no-watermark nonlinear, and fold-shuffled controls are consistent with interpreting the gain as score geometry, not as evidence of a strong stand-alone watermark signal.

\FloatBarrier

\section{Conclusion}

This paper asks what is lost when speech deepfake detection is reduced to one final score.
On the 4,080-utterance matched ASVspoof~5 intersection, retaining passive, probe, retrieval, and profile cues until late calibration improves score performance over fixed scalar fusion and keeps component fields visible.
The gain is concentrated, with most of the improvement arising from passive failure modes in A12 and A13.
Nonlinear controls leave unresolved whether the operational raw-score gap features cause the performance change.
In this setting, the keyed probe is a conditional, inspectable field, not an independent detector with a separately established predictive benefit.
Retrospective diagnostics show that the record can attach explicit fields to surfaced utterances, but they do not establish a deployment-time inspection policy.

\FloatBarrier

\bibliography{references}
\bibliographystyle{submission_style}

\appendix
\raggedbottom

\section{Additional Evidence}

The appendix collects the protocol details, passive-reference comparisons, boundary checks, and stress analyses that support the main text.

\subsection{Matched Subset and Operating Convention}

We document the examples, score orientation, and cost conventions used in the main text in Tables~\ref{tab:subset-coverage}--\ref{tab:cost-orientation}.

\begin{table}[H]
\caption{Coverage of the score subsets used in the paper. BF/S gives bona fide (BF) and spoof counts; source-tag counts use broad corpus labels. Duration is computed from referenced audio files when available.}
\label{tab:subset-coverage}
\centering
\fontsize{6.8}{7.8}\selectfont
\setlength{\tabcolsep}{2.8pt}
\begin{tabular}{lrrrrrrl}
\toprule

{\bfseries\boldmath Subset} & {\bfseries\boldmath Examples} & {\bfseries\boldmath Utter.} & {\bfseries\boldmath BF/S} & {\bfseries\boldmath Spk.} & {\bfseries\boldmath Source tags} & {\bfseries\boldmath Families} & {\bfseries\boldmath Duration} \\
\midrule
Matched main & 4,080 & 4,080 & 3,215/865 & 705 & 1 & 8 & 8.01 h \\
Retrieval pre-join & 40,000 & 40,000 & 20,000/20,000 & n/a & n/a & 8 & not available \\
WavLM matched & 4,080 & 4,080 & 3,215/865 & 705 & 1 & 8 & 8.01 h \\
\bottomrule
\end{tabular}
\end{table}

\begin{table}[H]
\caption{Passive-score orientation analysis for the held-out synthesis-family protocol. The gap is the mean passive bona fide probability on bona fide examples minus that on spoof examples; negative gaps indicate anti-oriented scores. Inverted EER uses the deterministic score complement.}
\label{tab:passive-polarity}
\centering
\fontsize{6.8}{7.8}\selectfont
\setlength{\tabcolsep}{3.2pt}
\begin{tabular}{lrrrrrrc}
\toprule

{\bfseries\boldmath Family} & {\bfseries\boldmath Spoof} & {\bfseries\boldmath $\bar p_{\mathrm{BF}}^{\mathrm{bon}}$} & {\bfseries\boldmath $\bar p_{\mathrm{BF}}^{\mathrm{spoof}}$} & {\bfseries\boldmath Gap} & {\bfseries\boldmath \mbox{Passive EER (\%)~$\downarrow$}} & {\bfseries\boldmath \mbox{Inv. EER (\%)~$\downarrow$}} & {\bfseries\boldmath Call} \\
\midrule
A09 & 111 & 0.627 & 0.088 & +0.539 & 9.92 & 90.08 & original \\
A10 & 124 & 0.627 & 0.072 & +0.555 & 8.73 & 91.27 & original \\
A11 & 121 & 0.627 & 0.060 & +0.568 & 6.43 & 93.57 & original \\
A12 & 91 & 0.627 & 0.945 & -0.318 & 81.31 & 18.69 & anti-oriented \\
A13 & 100 & 0.627 & 0.522 & +0.105 & 53.00 & 47.00 & near chance \\
A14 & 102 & 0.627 & 0.062 & +0.565 & 5.65 & 94.35 & original \\
A15 & 116 & 0.627 & 0.285 & +0.342 & 23.27 & 76.73 & original \\
A16 & 100 & 0.627 & 0.077 & +0.550 & 7.22 & 92.78 & original \\
\bottomrule
\end{tabular}
\end{table}

\begin{table}[H]
\caption{Cost-orientation sensitivity for the 4,080 matched examples. minDCF$_{\mathrm{BF}}$ treats bona fide speech as the target class; minDCF$_{\mathrm{spoof}}$ inverts labels and scores. Values near 1.0 mark the default-cost ceiling.}
\label{tab:cost-orientation}
\centering
\fontsize{6.8}{7.8}\selectfont
\setlength{\tabcolsep}{3.0pt}
\begin{tabular}{llrrr}
\toprule

{\bfseries\boldmath Group} & {\bfseries\boldmath Score} & {\bfseries\boldmath \mbox{EER (\%)~$\downarrow$}} & {\bfseries\boldmath \mbox{minDCF$_{\mathrm{BF}}$~$\downarrow$}} & {\bfseries\boldmath \mbox{minDCF$_{\mathrm{spoof}}$~$\downarrow$}} \\
\midrule
Main ablations & Passive CNN & 21.15 & 1.0000 & 0.8579 \\
Main ablations & Watermark probe & 48.67 & 0.9972 & 0.9988 \\
Main ablations & Fixed record & 21.03 & 0.9984 & 0.9549 \\
Main ablations & Retrieval kNN & 15.84 & 0.9406 & 0.9988 \\
Main ablations & Fixed record + retrieval & 11.91 & 0.8283 & 0.6659 \\
Main ablations & Profile fusion & 12.96 & 0.7673 & 0.6974 \\
Main ablations & Distance-only control & 62.68 & 1.0000 & 1.0000 \\
Calibration & Fixed record + retrieval & 11.91 & 0.8283 & 0.6659 \\
Calibration & Cross-fit score fusion & 11.91 & 0.9107 & 0.7295 \\
Calibration & Cross-fit nonlinear control & 10.16 & 0.8986 & 0.7618 \\
Calibration & Cross-fit absolute-gap record & 8.43 & 0.9365 & 0.6913 \\
\bottomrule
\end{tabular}
\end{table}

\subsection{Passive-Detector Context}

These tables place the matched decision-record comparison beside stronger passive SSL references and separate external stress settings.
Appendix Table~\ref{tab:ssl-duration-context} shows that longer crops do not improve frozen SSL baselines on these development runs.
Appendix Table~\ref{tab:external-ssl-context} separates outside-corpus transfer from the matched comparison.
Appendix Tables~\ref{tab:matched-ssl-context} and~\ref{tab:wavlm-card-context} place the matched WavLM reference on the same 4,080 examples as the decision-record analysis.
Appendix Table~\ref{tab:self-vc-ssl-context} summarizes the self-voice-conversion stress condition.

\begin{table}[H]
\caption{Frozen-SSL passive baseline context on full ASVspoof 5 evaluation scores. These entries contextualize the archived CNN baseline and summarize detector strength on evaluation examples that differ from Table~\ref{tab:matched-results} and from the official challenge-submission pool.}
\label{tab:ssl-eval-context}
\centering
\fontsize{6.6}{7.5}\selectfont
\setlength{\tabcolsep}{0.5pt}
\begin{tabularx}{0.92\linewidth}{@{}>{\raggedright\arraybackslash}p{0.18\linewidth}*{6}{>{\centering\arraybackslash}X}@{}}
\toprule

{\bfseries\boldmath Backbone} & {\bfseries\boldmath Runs} & {\bfseries\boldmath \shortstack[c]{Train\\ cap}} & {\bfseries\boldmath \shortstack[c]{Eval\\ examples}} & {\bfseries\boldmath \shortstack[c]{Best EER\\(\%)~$\downarrow$}} & {\bfseries\boldmath \shortstack[c]{Median EER\\(\%)~$\downarrow$}} & {\bfseries\boldmath \shortstack[c]{EER range\\(\%)~$\downarrow$}} \\
\midrule
WavLM-large & 12 & 20,000 & 680,774 & 9.06 & 9.83 & 9.06--11.72 \\
\shortstack[l]{Wav2Vec2-large\\LV60} & 11 & 20,000 & 680,774 & 11.69 & 11.91 & 11.69--12.03 \\
HuBERT-large & 11 & 20,000 & 680,774 & 12.02 & 12.21 & 12.02--12.35 \\
\bottomrule
\end{tabularx}
\end{table}

\begin{table}[H]
\caption{Full-development SSL duration controls for passive detector context. Entries use 20k-per-class ASVspoof 5 Track 1 development runs with 40,000 scored examples. EER and $\Delta$Median are percentages; $\Delta$Median compares each crop with the same backbone's completed 4 s median.}
\label{tab:ssl-duration-context}
\centering
\fontsize{6.8}{7.8}\selectfont
\setlength{\tabcolsep}{3.0pt}
\begin{tabular}{lrrrrrr}
\toprule

{\bfseries\boldmath Backbone} & {\bfseries\boldmath Crop} & {\bfseries\boldmath Runs} & {\bfseries\boldmath \mbox{Best EER (\%)~$\downarrow$}} & {\bfseries\boldmath \mbox{Median EER (\%)~$\downarrow$}} & {\bfseries\boldmath \mbox{EER range (\%)~$\downarrow$}} & {\bfseries\boldmath $\Delta$Median} \\
\midrule
WavLM-large & 4 s & 12 & 7.00 & 8.57 & 7.00--11.16 & +0.00 \\
WavLM-large & 6 s & 8 & 9.03 & 9.70 & 9.03--10.87 & +1.13 \\
WavLM-large & 8 s & 5 & 11.15 & 12.80 & 11.15--14.92 & +4.22 \\
WavLM-large & 10 s & 8 & 10.91 & 13.17 & 10.91--15.04 & +4.60 \\
WavLM-large & 12 s & 4 & 13.00 & 13.54 & 13.00--13.93 & +4.97 \\
WavLM-large & 14 s & 4 & 11.29 & 11.64 & 11.29--12.35 & +3.07 \\
WavLM-large & 16 s & 4 & 8.67 & 9.03 & 8.67--9.32 & +0.45 \\
Wav2Vec2-large LV60 & 4 s & 11 & 9.21 & 9.45 & 9.21--9.73 & +0.00 \\
Wav2Vec2-large LV60 & 6 s & 7 & 9.04 & 9.34 & 9.04--9.83 & -0.11 \\
Wav2Vec2-large LV60 & 8 s & 4 & 9.84 & 10.13 & 9.84--10.44 & +0.68 \\
Wav2Vec2-large LV60 & 10 s & 7 & 9.80 & 10.20 & 9.80--10.88 & +0.74 \\
Wav2Vec2-large LV60 & 12 s & 4 & 11.27 & 11.43 & 11.27--12.62 & +1.98 \\
Wav2Vec2-large LV60 & 14 s & 4 & 10.96 & 11.34 & 10.96--12.03 & +1.89 \\
Wav2Vec2-large LV60 & 16 s & 4 & 10.48 & 11.02 & 10.48--11.20 & +1.57 \\
HuBERT-large & 4 s & 11 & 8.94 & 9.21 & 8.94--9.73 & +0.00 \\
HuBERT-large & 6 s & 7 & 9.45 & 10.04 & 9.45--10.64 & +0.83 \\
HuBERT-large & 8 s & 4 & 16.07 & 17.73 & 16.07--18.19 & +8.52 \\
HuBERT-large & 10 s & 7 & 19.07 & 19.98 & 19.07--21.00 & +10.78 \\
HuBERT-large & 12 s & 4 & 17.95 & 18.84 & 17.95--19.34 & +9.63 \\
HuBERT-large & 14 s & 4 & 19.57 & 20.17 & 19.57--21.34 & +10.96 \\
HuBERT-large & 16 s & 4 & 20.71 & 22.52 & 20.71--23.01 & +13.31 \\
\bottomrule
\end{tabular}
\end{table}

Training-size effects do not explain the decision-record comparison.
In Table~\ref{tab:ssl-data-scale-context}, Wav2Vec2-large LV60 and HuBERT-large improve modestly from 20k to 120k examples per class, while WavLM-large is worse at the larger caps than at 20k.
We use the selected WavLM setting only as a strong matched passive reference.

\begin{table}[H]
\caption{Full-development SSL training-size controls for passive detector context. Entries aggregate completed 4 s ASVspoof 5 Track 1 development runs by frozen SSL backbone and per-class training cap. EER and $\Delta$Median are percentages; $\Delta$Median compares each entry with the same backbone's completed 20k-per-class median.}
\label{tab:ssl-data-scale-context}
\centering
\fontsize{6.8}{7.8}\selectfont
\setlength{\tabcolsep}{2.8pt}
\begin{tabular}{lrrrrrr}
\toprule

{\bfseries\boldmath Backbone} & {\bfseries\boldmath Size/class} & {\bfseries\boldmath Examples} & {\bfseries\boldmath Runs} & {\bfseries\boldmath \mbox{Best EER (\%)~$\downarrow$}} & {\bfseries\boldmath \mbox{Median EER (\%)~$\downarrow$}} & {\bfseries\boldmath $\Delta$Median} \\
\midrule
WavLM-large & 20,000 & 40,000 & 12/12 & 7.00 & 8.57 & +0.00 \\
WavLM-large & 40,000 & 71,334 & 10/10 & 11.17 & 11.92 & +3.35 \\
WavLM-large & 80,000 & 111,334 & 10/10 & 9.58 & 10.52 & +1.94 \\
WavLM-large & 120,000 & 140,950 & 10/10 & 9.53 & 10.35 & +1.78 \\
Wav2Vec2-large LV60 & 20,000 & 40,000 & 11/11 & 9.21 & 9.45 & +0.00 \\
Wav2Vec2-large LV60 & 80,000 & 111,334 & 10/10 & 8.92 & 9.02 & -0.43 \\
Wav2Vec2-large LV60 & 120,000 & 140,950 & 10/10 & 8.89 & 8.94 & -0.51 \\
HuBERT-large & 20,000 & 40,000 & 11/11 & 8.94 & 9.21 & +0.00 \\
HuBERT-large & 80,000 & 111,334 & 10/10 & 8.37 & 8.65 & -0.55 \\
HuBERT-large & 120,000 & 140,950 & 10/10 & 8.41 & 8.55 & -0.66 \\
\bottomrule
\end{tabular}
\end{table}

\begin{table}[H]
\caption{Frozen SSL outside-corpus results. Each entry summarizes the same 60 ASVspoof-5-trained SSL scoring runs spanning three encoders, two training-set sizes, and ten seeds on one outside evaluation set. The ASVspoof 2021 DeepFake (DF) subset is named explicitly in the rows. The table provides passive-detector context beyond the main ASVspoof 5 comparison.}
\label{tab:external-ssl-context}
\centering
\fontsize{6.8}{7.8}\selectfont
\setlength{\tabcolsep}{1.0pt}
\begin{tabularx}{1.0\linewidth}{@{}>{\raggedright\arraybackslash}p{0.18\linewidth}*{6}{>{\centering\arraybackslash}X}@{}}
\toprule

{\bfseries\boldmath External set} & {\bfseries\boldmath Runs} & {\bfseries\boldmath \shortstack[c]{Examples\\per run}} & {\bfseries\boldmath \shortstack[c]{Best EER\\(\%)~$\downarrow$}} & {\bfseries\boldmath \shortstack[c]{Median EER\\(\%)~$\downarrow$}} & {\bfseries\boldmath \shortstack[c]{Worst EER\\(\%)~$\downarrow$}} & {\bfseries\boldmath \shortstack[c]{Median ECE~$\downarrow$}} \\
\midrule
In-The-Wild & 60/60 & 31,779 & 14.25 & 24.60 & 28.33 & 0.2581 \\
WaveFake & 60/60 & 147,366 & 34.53 & 40.29 & 81.17 & 0.8528 \\
ASVspoof 2021 DF & 60/60 & 611,829 & 15.99 & 22.33 & 28.21 & 0.6046 \\
\bottomrule
\end{tabularx}
\end{table}

\begin{table}[H]
\caption{Matched frozen SSL passive context on the same 4,080 examples as Table~\ref{tab:matched-results}. Runs vary backbone, seed, and training size. The best entry is chosen within each backbone; no entry includes watermark, retrieval, or profile evidence.}
\label{tab:matched-ssl-context}
\centering
\fontsize{6.8}{7.8}\selectfont
\setlength{\tabcolsep}{1.0pt}
\begin{tabularx}{1.0\linewidth}{@{}>{\raggedright\arraybackslash}p{0.18\linewidth}*{6}{>{\centering\arraybackslash}X}@{}}
\toprule

{\bfseries\boldmath Backbone} & {\bfseries\boldmath Runs} & {\bfseries\boldmath \shortstack[c]{Examples\\per run}} & {\bfseries\boldmath \shortstack[c]{Best EER\\(\%)~$\downarrow$}} & {\bfseries\boldmath \shortstack[c]{Median EER\\(\%)~$\downarrow$}} & {\bfseries\boldmath \shortstack[c]{EER range\\(\%)~$\downarrow$}} & {\bfseries\boldmath \shortstack[c]{minDCF /\\ ECE~$\downarrow$}} \\
\midrule
WavLM-large & 22/22 & 4,080 & 6.71 & 9.60 & 6.71--13.63 & 0.9776/0.0453 \\
HuBERT-large & 11/11 & 4,080 & 8.21 & 9.24 & 8.21--9.48 & 0.9792/0.0416 \\
Wav2Vec2-large & 11/11 & 4,080 & 9.14 & 9.83 & 9.14--10.05 & 0.7801/0.0318 \\
\bottomrule
\end{tabularx}
\end{table}

\begin{table}[H]
\caption{WavLM passive context on the same 4,080 matched examples as Table~\ref{tab:matched-results}. The passive entry is the lowest-EER WavLM-large run among 22 matched passive SSL runs. The sweep median is 9.60\% EER and serves as the scalar baseline for comparison. W, R, and P denote watermark, retrieval, and profile fields.}
\label{tab:wavlm-card-context}
\vspace{3pt}
\centering
\fontsize{6.20}{7.00}\selectfont
\setlength{\tabcolsep}{0.8pt}
\renewcommand{\arraystretch}{1.07}
\begin{tabularx}{1.0\linewidth}{@{}>{\raggedright\arraybackslash}p{0.27\linewidth}>{\raggedright\arraybackslash}p{0.10\linewidth}*{6}{>{\centering\arraybackslash}X}@{}}
\toprule
{\bfseries\boldmath System} & {\bfseries\boldmath Fit} & {\bfseries\boldmath \shortstack[c]{Pooled\\EER (\%)~$\downarrow$}} & {\bfseries\boldmath \shortstack[c]{Family\\EER (\%)~$\downarrow$}} & {\bfseries\boldmath \shortstack[c]{Fold\\EER (\%)~$\downarrow$}} & {\bfseries\boldmath \shortstack[c]{$\Delta$EER\\(\%)~$\downarrow$}} & {\bfseries\boldmath \shortstack[c]{minDCF$_{\mathrm{BF}}$\\$\downarrow$}} & {\bfseries\boldmath ECE~$\downarrow$} \\
\midrule
WavLM passive (best run) & passive & 6.71 & 5.33 & 5.00 & +0.00 & 0.9776 & 0.0453 \\
Retrieval kNN & no fit & 15.84 & 13.85 & 13.69 & +9.12 & 0.9406 & 0.1451 \\
WavLM + watermark & fixed & 10.86 & 8.82 & -- & +4.15 & 0.9919 & 0.2813 \\
WavLM + watermark + retrieval & fixed & 11.33 & 9.53 & 8.80 & +4.61 & 0.9872 & 0.2019 \\
WavLM + watermark + retrieval + profile & fixed & 10.17 & 8.09 & 7.83 & +3.46 & 0.9708 & 0.2861 \\
\textbf{WavLM decision record} & \textbf{learned} & \textbf{9.14} & \textbf{--} & \textbf{6.56} & \textbf{+2.43} & \textbf{0.9984} & \textbf{0.0657} \\
WavLM passive (cross-fit) & passive fit & 11.79 & -- & 5.00 & +5.08 & 0.9823 & 0.1076 \\
WavLM score fusion & learned scalar & 9.72 & -- & 6.85 & +3.01 & 0.9991 & 0.0712 \\
\bottomrule
\end{tabularx}
\renewcommand{\arraystretch}{1.0}
\end{table}

\begin{table}[H]
\caption{Frozen-SSL self-voice-conversion stress summary. Each entry aggregates completed runs on the same 2,000-example converted-speech condition used for the passive CNN entry in Table~\ref{tab:stress-diagnostics}. Ranges show run-to-run spread within a model family; EER values are percentages.}
\label{tab:self-vc-ssl-context}
\centering
\fontsize{6.8}{7.8}\selectfont
\setlength{\tabcolsep}{4.0pt}
\begin{tabular}{lrrrrr}
\toprule

{\bfseries\boldmath Model family} & {\bfseries\boldmath Runs} & {\bfseries\boldmath \mbox{EER range (\%)~$\downarrow$}} & {\bfseries\boldmath \mbox{Mean EER (\%)~$\downarrow$}} & {\bfseries\boldmath \mbox{minDCF range~$\downarrow$}} & {\bfseries\boldmath \mbox{Mean ECE~$\downarrow$}} \\
\midrule
HuBERT-large & 21 & 25.60--34.50 & 29.89 & 0.9480--1.0000 & 0.4564 \\
Wav2Vec2-large & 26 & 23.70--36.50 & 29.65 & 0.9890--1.0000 & 0.4331 \\
WavLM-base-plus & 12 & 27.30--30.00 & 28.21 & 0.9990--1.0000 & 0.4431 \\
WavLM-large & 34 & 32.90--43.40 & 37.25 & 0.9970--1.0000 & 0.4700 \\
\bottomrule
\end{tabular}
\end{table}

\subsection{Probe and Retrieval Boundary Checks}

These tables show what the probe, retrieval, and profile streams support under the available metadata and controls.

\begin{table}[H]
\caption{Score-level watermark-availability sensitivity on the 4,080 matched examples. Stored watermark columns are removed, neutralized, or inverted while other scores stay fixed. The table separates probe availability from failure-mode behavior.}
\label{tab:watermark-availability}
\vspace{3pt}
\centering
\fontsize{6.25}{7.00}\selectfont
\setlength{\tabcolsep}{0.8pt}
\renewcommand{\arraystretch}{1.10}
\begin{tabularx}{1.0\linewidth}{@{}>{\raggedright\arraybackslash}p{0.27\linewidth}*{6}{>{\centering\arraybackslash}X}>{\raggedright\arraybackslash}p{0.10\linewidth}@{}}
\toprule
{\bfseries\boldmath Scenario} & {\bfseries\boldmath \mbox{W field}} & {\bfseries\boldmath \shortstack[c]{Pooled\\EER (\%)~$\downarrow$}} & {\bfseries\boldmath \shortstack[c]{Family\\EER (\%)~$\downarrow$}} & {\bfseries\boldmath \shortstack[c]{minDCF$_{\mathrm{BF}}$\\$\downarrow$}} & {\bfseries\boldmath ECE~$\downarrow$} & {\bfseries\boldmath Brier~$\downarrow$} & {\bfseries\boldmath \shortstack[c]{Worst\\family~$\downarrow$}} \\
\midrule
Full record: passive, probe, retrieval & yes & 11.91 & 10.27 & 0.8283 & 0.2609 & 0.1415 & \mbox{A15 (25.85)} \\
Probe unavailable: reweighted passive/retrieval & no & 12.46 & 9.35 & 0.7692 & 0.1878 & 0.1098 & \mbox{A15 (21.69)} \\
Probe unavailable: neutral slot & no & 12.46 & 9.35 & 0.7692 & 0.2582 & 0.1375 & \mbox{A15 (21.69)} \\
Inverted probe & yes & 12.37 & 9.51 & 0.7552 & 0.2095 & 0.1130 & \mbox{A15 (22.40)} \\
Full record + profile (all streams) & yes & 12.96 & 9.86 & 0.7673 & 0.3270 & 0.1834 & \mbox{A15 (22.33)} \\
Probe unavailable + profile & no & 12.72 & 9.84 & 0.7496 & 0.2522 & 0.1366 & \mbox{A15 (22.42)} \\
\bottomrule
\end{tabularx}
\renewcommand{\arraystretch}{1.0}
\end{table}

The raw keyed-probe statistic, threshold, and decision counts used to interpret the probe channel are reported in Appendix Table~\ref{tab:watermark-operating-counts}.

\begin{table}[H]
\caption{Operating counts for watermark-probe conditions. Probe mean, probe standard deviation (SD), and threshold come from the raw keyed-probe statistic; scores at or above the threshold are accepted as marked. Direct pairs use marked and unmarked pairs, and the remaining conditions come from the conversion study.}
\label{tab:watermark-operating-counts}
\centering
\fontsize{6.8}{7.6}\selectfont
\setlength{\tabcolsep}{1.0pt}
\renewcommand{\arraystretch}{1.05}
\begin{tabularx}{1.0\linewidth}{@{}>{\raggedright\arraybackslash}p{0.17\linewidth}>{\centering\arraybackslash}p{0.055\linewidth}>{\centering\arraybackslash}p{0.085\linewidth}>{\centering\arraybackslash}p{0.078\linewidth}>{\centering\arraybackslash}p{0.080\linewidth}>{\centering\arraybackslash}p{0.10\linewidth}>{\centering\arraybackslash}p{0.10\linewidth}>{\raggedright\arraybackslash}X@{}}
\toprule
{\bfseries\boldmath Condition} & {\bfseries\boldmath $n$} & {\bfseries\boldmath \shortstack[c]{Probe\\mean}} & {\bfseries\boldmath \shortstack[c]{Probe\\SD}} & {\bfseries\boldmath \shortstack[c]{Probe\\threshold}} & {\bfseries\boldmath \shortstack[c]{Marked\\acc./rej.}} & {\bfseries\boldmath \shortstack[c]{Unmarked\\acc./rej.}} & {\bfseries\boldmath Interpretation} \\
\midrule
Probe wiring condition & 20,000 & 0.01258 & 0.01261 & 0.01421 & 10,000/0 & 0/10,000 & direct-pair positive control \\
Source-pair control & 2,000 & 0.00007 & 0.00099 & -0.00481 & 1,000/0 & 1,000/0 & threshold accepts both classes \\
Converted-pair condition & 2,000 & 0.00000 & 0.00086 & -0.00085 & 905/95 & 896/104 & weak mark readout after conversion \\
\bottomrule
\end{tabularx}
\renewcommand{\arraystretch}{1.0}
\end{table}

\begin{table}[H]
\caption{Field-semantics and applicability boundary for decision-record fields. The table states how each channel is interpreted in this paper and distinguishes keyed-probe behavior from authentication of an original utterance.}
\label{tab:provenance-threat-model}
\centering
\fontsize{6.8}{7.8}\selectfont
\setlength{\tabcolsep}{2.6pt}
\renewcommand{\arraystretch}{1.06}
\begin{tabular}{@{}>{\raggedright\arraybackslash}p{0.20\linewidth}>{\raggedright\arraybackslash}p{0.18\linewidth}>{\raggedright\arraybackslash}p{0.34\linewidth}>{\raggedright\arraybackslash}p{0.20\linewidth}@{}}
\toprule

{\bfseries\boldmath Scenario} & {\bfseries\boldmath Field(s) used} & {\bfseries\boldmath Reading in this paper} & {\bfseries\boldmath Open requirement} \\
\midrule
Known key, directly marked signal & $s_w$ on marked/unmarked pairs & probe wiring and operating threshold under the recorded key & performance on unmarked source audio \\
Derived marked copy in matched examples & $s_p,s_w,s_r,s_m$ and gaps & keyed-probe response and cross-stream raw-score gaps under the same derived condition for both classes & proof that the original benchmark file carried a mark \\
No key or absent mark & $s_p,s_r,s_m$ plus availability status & passive, retrieval, and profile evidence stays available without the probe field & watermark authentication \\
Removal, corruption, or conversion & probe scores under stress & near-chance scores identify a failure condition for the probe channel & mark survival under attack \\
Retrieved/profile support available & $s_r,s_m,c_r$ and neighbor metadata & support-set context after family exclusion and overlap analysis & prompt, channel, or collection independence \\
\bottomrule
\end{tabular}
\renewcommand{\arraystretch}{1.0}
\end{table}

\begin{table}[H]
\caption{Retrieval subset selection and nearest-neighbor overlap analysis. Full scores precede the matched fusion join, and matched examples are used in Table~\ref{tab:matched-results}. The final columns report held-out-family leakage in the nearest neighbor and top-$k$ list; coarse source tags describe corpus-level grouping and do not identify prompt or recording-source separation.}
\label{tab:retrieval-subset}
\centering
\fontsize{6.5}{7.4}\selectfont
\setlength{\tabcolsep}{1.0pt}
\begin{tabularx}{0.96\linewidth}{@{}>{\raggedright\arraybackslash}p{0.07\linewidth}*{7}{>{\centering\arraybackslash}X}@{}}
\toprule

{\bfseries\boldmath Family} & {\bfseries\boldmath \shortstack[c]{Full\\ spoof}} & {\bfseries\boldmath \shortstack[c]{Matched\\ spoof}} & {\bfseries\boldmath Retained} & {\bfseries\boldmath \shortstack[c]{Full EER\\(\%)~$\downarrow$}} & {\bfseries\boldmath \shortstack[c]{Matched\\ EER (\%)~$\downarrow$}} & {\bfseries\boldmath \shortstack[c]{Same-family\\NN}} & {\bfseries\boldmath \shortstack[c]{Same-family\\top-$k$}} \\
\midrule
A09 & 2,479 & 111 & 4.5\% & 15.36 & 13.51 & \shortstack[c]{0/2,479;\\0/111} & 0/2,479 \\
A10 & 2,548 & 124 & 4.9\% & 14.84 & 15.22 & \shortstack[c]{0/2,548;\\0/124} & 0/2,548 \\
A11 & 2,568 & 121 & 4.7\% & 26.05 & 24.65 & \shortstack[c]{0/2,568;\\0/121} & 0/2,568 \\
A12 & 2,457 & 91 & 3.7\% & 1.96 & 2.03 & \shortstack[c]{0/2,457;\\0/91} & 0/2,457 \\
A13 & 2,526 & 100 & 4.0\% & 23.48 & 19.19 & \shortstack[c]{0/2,526;\\0/100} & 0/2,526 \\
A14 & 2,450 & 102 & 4.2\% & 1.95 & 1.91 & \shortstack[c]{0/2,450;\\0/102} & 0/2,450 \\
A15 & 2,426 & 116 & 4.8\% & 27.41 & 29.31 & \shortstack[c]{0/2,426;\\0/116} & 0/2,426 \\
A16 & 2,546 & 100 & 3.9\% & 6.48 & 5.00 & \shortstack[c]{0/2,546;\\0/100} & 0/2,546 \\
\bottomrule
\end{tabularx}
\end{table}

Subset matching changes the retrieval-score distribution relative to the full score set.
Appendix Table~\ref{tab:retrieval-join-selection} quantifies the shift.

\begin{table}[H]
\caption{Matched-subset sensitivity analysis for retrieval scores. The bootstrap column draws the same number of bona fide and spoof examples as the matched subset from the full retrieval score set. The final columns report the EER shift from that random-retention reference, the retained-minus-omitted shift in spoof-example bona fide scores, and the corresponding KS distance. Larger spoof-example scores indicate harder retrieval cases.}
\label{tab:retrieval-join-selection}
\centering
\vspace{3pt}
\fontsize{6.5}{7.2}\selectfont
\setlength{\tabcolsep}{2.5pt}
\renewcommand{\arraystretch}{1.04}
\begin{tabular}{@{}lrrrrrr@{}}
\toprule
{\bfseries\boldmath Family} & {\bfseries\boldmath Retained} & {\bfseries\boldmath \shortstack[c]{Match\\EER (\%)~$\downarrow$}} & {\bfseries\boldmath \shortstack[c]{Boot.\\EER (\%)~$\downarrow$}} & {\bfseries\boldmath \shortstack[c]{Match--\\rand. (\%)}} & {\bfseries\boldmath \shortstack[c]{Spoof\\shift~$\downarrow$}} & {\bfseries\boldmath \mbox{KS~$\downarrow$}} \\
\midrule
A09 & 4.5\% & 13.51 & 15.31 [12.60, 18.01] & -1.80 & -2.01 & 0.085 \\
A10 & 4.9\% & 15.22 & 14.77 [12.10, 17.58] & +0.45 & +0.21 & 0.058 \\
A11 & 4.7\% & 24.65 & 26.09 [22.11, 30.58] & -1.44 & -3.12 & 0.078 \\
A12 & 3.7\% & 2.03 & 2.23 [1.34, 3.30] & -0.19 & -0.24 & 0.017 \\
A13 & 4.0\% & 19.19 & 23.49 [19.00, 28.00] & -4.30 & -7.45 & 0.133 \\
A14 & 4.2\% & 1.91 & 1.96 [1.74, 2.18] & -0.05 & +0.07 & 0.009 \\
A15 & 4.8\% & 29.31 & 27.49 [23.07, 32.62] & +1.81 & +0.00 & 0.083 \\
A16 & 3.9\% & 5.00 & 6.27 [4.01, 8.99] & -1.26 & -1.80 & 0.046 \\
\bottomrule
\end{tabular}
\renewcommand{\arraystretch}{1.0}
\end{table}

The retrieval-neighbor audit in Appendix Table~\ref{tab:retrieval-neighbor-audit} complements the setup in Table~\ref{tab:protocol-details}.
Under the available metadata, it rules out exact family, utterance, speaker-ID, and audio-path reuse.
Broader source and channel overlap remain limitations.

\begin{table}[H]
\caption{Identifier-overlap analysis for retrieval/profile nearest neighbors. Counts come from the pre-join 40,000-example retrieval setting with held-out synthesis-family exclusion for spoof examples. The coarse source tag is corpus-level metadata; prompt, channel, and recording-source independence remain unresolved at finer granularity.}
\label{tab:retrieval-neighbor-audit}
\centering
\fontsize{6.8}{7.8}\selectfont
\setlength{\tabcolsep}{3.0pt}
\begin{tabular}{@{}>{\raggedright\arraybackslash}p{0.42\linewidth}rr>{\raggedright\arraybackslash}p{0.24\linewidth}@{}}
\toprule

{\bfseries\boldmath Identifier relation} & {\bfseries\boldmath Examples} & {\bfseries\boldmath Matches} & {\bfseries\boldmath Interpretation} \\
\midrule
Held-out family as nearest neighbor & 20,000 & 0/20,000 & blocked by family exclusion \\
Held-out family anywhere in top-k & 20,000 & 0/20,000 & blocked by family exclusion \\
Same utterance ID anywhere in top-k & 40,000 & 0/40,000 & not observed \\
Same speaker ID anywhere in top-k & 40,000 & 0/40,000 & not observed \\
Same audio path anywhere in top-k & 40,000 & 0/40,000 & not observed \\
Same coarse source tag anywhere in top-k & 40,000 & 40,000/40,000 & shared corpus-level tag \\
\bottomrule
\end{tabular}
\end{table}

\subsection{Family, Calibration, and Stress Evidence}

The remaining appendix material addresses three points: where pooled gains concentrate, how calibration terms behave, and why the active-selection and attack-stress studies remain outside the central matched-subset result.

\begin{table}[H]
\caption{Held-out synthesis-family uncertainty for Figure~\ref{fig:family-eer}. Counts give bona fide and spoof examples. EER is receiver operating characteristic (ROC)-interpolated and can differ from integer error-count ratios in small families. $\Delta_P$ and $\Delta_R$ compare the fixed retrieval-augmented rule with passive CNN and retrieval kNN. Bold intervals favor the retrieval-augmented rule; \textdagger{} marks intervals against it.}
\label{tab:families}
\centering
\fontsize{6.8}{7.8}\selectfont
\setlength{\tabcolsep}{2.6pt}
\begin{tabular}{lrrrrrrll}
\toprule

{\bfseries\boldmath Family} & {\bfseries\boldmath Bona.} & {\bfseries\boldmath Spoof} & {\bfseries\boldmath Passive} & {\bfseries\boldmath Ret.} & {\bfseries\boldmath Record} & {\bfseries\boldmath Profile} & {\bfseries\boldmath $\Delta_P$} & {\bfseries\boldmath $\Delta_R$} \\
\midrule
A09 & 3215 & 111 & 9.92 & 13.51 & 9.03 & \textbf{6.31} & -0.9 [-4.3, +3.6] & \textbf{-4.5 [-6.3, -2.0]} \\
A10 & 3215 & 124 & 8.73 & 15.22 & 8.09 & \textbf{5.64} & -0.6 [-3.2, +2.7] & \textbf{-7.1 [-8.2, -4.7]} \\
A11 & 3215 & 121 & \textbf{6.43} & 24.65 & 16.51 & 13.25 & +10.1 [+5.9, +15.8]\textdagger & \textbf{-8.1 [-10.8, -5.0]} \\
A12 & 3215 & 91 & 81.31 & \textbf{2.03} & 5.47 & 13.37 & \textbf{-75.8 [-79.2, -72.5]} & +3.4 [+2.5, +4.4]\textdagger \\
A13 & 3215 & 100 & 53.00 & 19.19 & 15.00 & \textbf{14.98} & \textbf{-38.0 [-48.1, -25.9]} & \textbf{-4.2 [-8.0, -1.0]} \\
A14 & 3215 & 102 & 5.65 & 1.91 & \textbf{0.22} & 0.97 & \textbf{-5.4 [-6.7, -3.9]} & \textbf{-1.7 [-2.9, -1.0]} \\
A15 & 3215 & 116 & 23.27 & 29.31 & 25.85 & \textbf{22.33} & +2.6 [-3.6, +8.0] & \textbf{-3.5 [-6.9, -0.2]} \\
A16 & 3215 & 100 & 7.22 & 5.00 & \textbf{2.00} & \textbf{2.00} & \textbf{-5.2 [-9.0, -2.0]} & \textbf{-3.0 [-4.6, -1.0]} \\
\bottomrule
\end{tabular}
\end{table}

\begin{figure}[t]
\centering
\begin{minipage}[t]{0.97\linewidth}
\centering
\begin{minipage}[t]{0.475\linewidth}
\centering
\makebox[\linewidth][c]{%
\begin{tikzpicture}
\begin{axis}[
  drawAxis,
  width=0.92\linewidth,
  height=4.55cm,
  scale only axis,
  xmin=0.65,
  xmax=8.35,
  ymin=0,
  ymax=0.32,
  title={(a) Retrieval and record variants},
  title style={font=\small, color=drawInk, yshift=-0.4ex},
  xlabel={held-out family},
  ylabel={EER (\%)},
  xtick={1,2,3,4,5,6,7,8},
  xticklabels={A09,A10,A11,A12,A13,A14,A15,A16},
  xticklabel style={font=\tiny, rotate=40, anchor=east},
  ytick={0,0.1,0.2,0.3},
  yticklabels={0,10,20,30},
  ymajorgrids=true,
  xmajorgrids=false,
  clip=false,
  tick align=outside,
  xlabel style={font=\small, yshift=0.2ex},
  ylabel style={font=\small, yshift=-0.4ex},
  tick label style={font=\scriptsize},
  every axis plot/.append style={line join=round, line cap=round},
  legend style={
    at={(0.50,1.03)},
    anchor=south,
    legend columns=2,
    draw=drawGrid!70,
    fill=white,
    rounded corners=1pt,
    font=\scriptsize,
    inner xsep=2.0pt,
    inner ysep=0.9pt,
    cells={anchor=west},
    /tikz/every even column/.append style={column sep=0.38em},
  },
  legend image post style={line width=0.60pt,mark size=0.90pt},
]
\addplot+[line width=0.68pt, dash pattern=on 4.4pt off 2.0pt, mark=square*, mark size=1.10pt, draw=plotGreen!82!black, mark options={fill=white, draw=plotGreen!82!black, line width=0.18pt}] table[x=idx,y=retrieval,col sep=comma] {figures/family_eer.dat};
\addlegendentry{Retrieval}
\addplot+[line width=0.72pt, solid, mark=triangle*, mark size=1.14pt, draw=plotBlue!88!black, mark options={fill=plotBlue!72!white, draw=white, line width=0.18pt}] table[x=idx,y=draw,col sep=comma] {figures/family_eer.dat};
\addlegendentry{Fixed retrieval rule}
\addplot+[line width=0.64pt, densely dotted, mark=diamond*, mark size=1.08pt, draw=drawPlum!82!black, mark options={fill=white, draw=drawPlum!82!black, line width=0.18pt}] table[x=idx,y=profile,col sep=comma] {figures/family_eer.dat};
\addlegendentry{Fixed retrieval + profile}
\end{axis}
\end{tikzpicture}%
}
\end{minipage}\hfill%
\begin{minipage}[t]{0.475\linewidth}
\centering
\makebox[\linewidth][c]{%
\begin{tikzpicture}
\begin{axis}[
  drawAxis,
  width=0.92\linewidth,
  height=4.55cm,
  scale only axis,
  xmin=0.65,
  xmax=8.35,
  ymin=0,
  ymax=0.86,
  title={(b) Passive CNN, full scale},
  title style={font=\small, color=drawInk, yshift=-0.4ex},
  xlabel={held-out family},
  ylabel={EER (\%)},
  xtick={1,2,3,4,5,6,7,8},
  xticklabels={A09,A10,A11,A12,A13,A14,A15,A16},
  xticklabel style={font=\tiny, rotate=40, anchor=east},
  ytick={0,0.25,0.50,0.75},
  yticklabels={0,25,50,75},
  xmajorgrids=false,
  ymajorgrids=true,
  tick align=outside,
  xlabel style={font=\small, yshift=0.2ex},
  ylabel style={font=\small, yshift=-0.4ex},
  tick label style={font=\scriptsize},
  every axis plot/.append style={line join=round, line cap=round},
]
\addplot+[line width=0.72pt, mark=*, mark size=1.12pt, draw=drawSlate!88, mark options={fill=drawSlate!70, draw=white, line width=0.18pt}] table[x=idx,y=passive,col sep=comma] {figures/family_eer.dat};
\node[anchor=east,font=\scriptsize,text=drawSlate,fill=drawPaper,fill opacity=0.97,text opacity=1,inner sep=1.2pt,rounded corners=0.9pt] at (axis cs:3.78,0.812) {A12: 81.3};
\node[anchor=west,font=\scriptsize,text=drawSlate,fill=drawPaper,fill opacity=0.97,text opacity=1,inner sep=1.2pt,rounded corners=0.9pt] at (axis cs:5.20,0.530) {A13: 53.0};
\end{axis}
\end{tikzpicture}%
}
\end{minipage}%
\end{minipage}
\vspace{1.2ex}
\caption{Held-out EER by synthesis family. Panel (a) zooms the retrieval/record variants; panel (b) keeps the passive-CNN failures on the full scale. Appendix Table~\ref{tab:families} gives exact values and paired intervals.}
\label{fig:family-eer}
\end{figure}
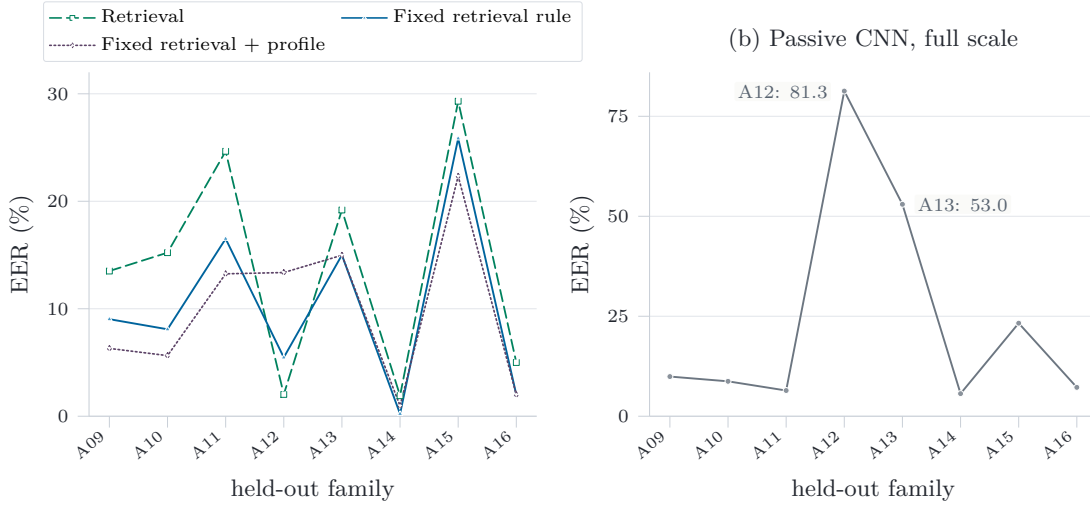

Table~\ref{tab:families} shows that the pooled gain is concentrated in a small subset of families; Appendix Figure~\ref{fig:family-eer} makes the pattern easier to see.
Appendix Table~\ref{tab:family-influence} confirms that excluding A12 and A13 largely removes the pooled advantage.
Appendix Table~\ref{tab:passive-card-error-accounting} shows that most of the net reduction comes from bona fide corrections in those families.

\begin{table}[H]
\caption{Family-influence sensitivity for the matched comparison. Each entry recomputes family-macro EER after excluding the named held-out spoof family or family pair from Table~\ref{tab:families}. The deltas compare the fixed retrieval-augmented rule with passive CNN and retrieval kNN under the same exclusion.}
\label{tab:family-influence}
\centering
\fontsize{6.8}{7.8}\selectfont
\setlength{\tabcolsep}{3.0pt}
\begin{tabular}{lrrrrrr}
\toprule

{\bfseries\boldmath Excluded} & {\bfseries\boldmath Passive} & {\bfseries\boldmath Ret.} & {\bfseries\boldmath Record} & {\bfseries\boldmath Profile} & {\bfseries\boldmath $\Delta_P$} & {\bfseries\boldmath $\Delta_R$} \\
\midrule
A09 & 26.52 & 13.90 & 10.45 & 10.36 & -16.07 & -3.45 \\
A10 & 26.69 & 13.66 & 10.58 & 10.46 & -16.10 & -3.07 \\
A11 & 27.01 & 12.31 & 9.38 & 9.37 & -17.63 & -2.93 \\
A12 & 16.32 & 15.54 & 10.96 & 9.35 & -5.36 & -4.58 \\
A13 & 20.36 & 13.09 & 9.60 & 9.12 & -10.77 & -3.49 \\
A14 & 27.13 & 15.56 & 11.71 & 11.13 & -15.42 & -3.85 \\
A15 & 24.61 & 11.64 & 8.05 & 8.07 & -16.56 & -3.60 \\
A16 & 26.90 & 15.12 & 11.45 & 10.98 & -15.45 & -3.66 \\
A12+A13 & 10.20 & 14.93 & 10.28 & 8.42 & 0.08 & -4.65 \\
\bottomrule
\end{tabular}
\end{table}

\begin{table}[H]
\caption{Threshold-error accounting for the pooled passive-CNN versus fixed retrieval-augmented rule comparison. Both systems use their own pooled EER threshold. Corrected examples are passive-CNN errors that the retrieval-augmented rule classifies correctly; new errors are errors made by the retrieval-augmented rule among passive-CNN correct decisions. Share divides the net error reduction by the full-set total.}
\label{tab:passive-card-error-accounting}
\centering
\fontsize{6.8}{7.8}\selectfont
\setlength{\tabcolsep}{2.4pt}
\begin{tabular}{lrrrrrrr}
\toprule

{\bfseries\boldmath Group} & {\bfseries\boldmath Examples} & {\bfseries\boldmath Passive err.} & {\bfseries\boldmath Fixed-rule err.} & {\bfseries\boldmath Corrected} & {\bfseries\boldmath New err.} & {\bfseries\boldmath Net} & {\bfseries\boldmath Share (\%)} \\
\midrule
All examples & 4,080 & 863 & 486 & 707 & 330 & +377 & +100.0 \\
Bona fide & 3,215 & 680 & 383 & 539 & 242 & +297 & +78.8 \\
A09 & 111 & 5 & 8 & 5 & 8 & -3 & -0.8 \\
A10 & 124 & 2 & 6 & 2 & 6 & -4 & -1.1 \\
A11 & 121 & 0 & 30 & 0 & 30 & -30 & -8.0 \\
A12 & 91 & 90 & 0 & 90 & 0 & +90 & +23.9 \\
A13 & 100 & 55 & 17 & 54 & 16 & +38 & +10.1 \\
A14 & 102 & 0 & 0 & 0 & 0 & +0 & +0.0 \\
A15 & 116 & 30 & 42 & 16 & 28 & -12 & -3.2 \\
A16 & 100 & 1 & 0 & 1 & 0 & +1 & +0.3 \\
\bottomrule
\end{tabular}
\end{table}

\begin{table}[H]
\caption{Retrospective selective risk on matched examples. Each system uses its own EER threshold and defers the nearest 10\% in rank space. The area under the risk--coverage curve (AURC) averages retained risk over accepted-prefix coverages and is given with a 95\% confidence interval (CI).}
\label{tab:evidence-card-triage}
\centering
\fontsize{6.6}{7.5}\selectfont
\setlength{\tabcolsep}{0.5pt}
\begin{tabularx}{\linewidth}{@{}>{\raggedright\arraybackslash}p{0.30\linewidth}*{6}{>{\centering\arraybackslash}X}@{}}
\toprule

{\bfseries\boldmath System} & {\bfseries\boldmath Errors} & {\bfseries\boldmath \shortstack{Decision\\error\\(\%)~$\downarrow$}} & {\bfseries\boldmath \shortstack{Error\\capture at\\10\%\\(\%)~$\uparrow$}} & {\bfseries\boldmath \shortstack{Review\\precision\\(\%)~$\uparrow$}} & {\bfseries\boldmath \shortstack{Retained\\error\\(\%)~$\downarrow$}} & {\bfseries\boldmath \shortstack{AURC\\(\%)~$\downarrow$\\(95\% CI)}} \\
\midrule
Passive CNN & 863/4,080 & 21.15 & 22.83 & 48.28 & 18.14 & \shortstack[c]{14.85\\(13.24--16.47)} \\
Retrieval kNN & 646/4,080 & 15.83 & 29.26 & 46.32 & 12.45 & \shortstack[c]{5.74\\(5.02--6.50)} \\
Fixed retrieval-augmented rule & 486/4,080 & 11.91 & 38.89 & 46.32 & 8.09 & \shortstack[c]{2.49\\(2.15--2.84)} \\
\rowcolor{blue!6}
\textbf{Proposed decision record} & 344/4,080 & 8.43 & 47.97 & 40.44 & 4.87 & \shortstack[c]{1.79\\(1.45--2.16)} \\
WavLM passive (best run) & 274/4,080 & 6.72 & 60.58 & 40.69 & 2.94 & \shortstack[c]{1.14\\(0.84--1.50)} \\
WavLM decision record & 373/4,080 & 9.14 & 50.40 & 46.08 & 5.04 & \shortstack[c]{2.96\\(2.25--3.74)} \\
\bottomrule
\end{tabularx}
\end{table}

\begin{table}[H]
\caption{Retrospective diagnostic-cue analysis for the cross-fit decision-record model. Cues follow the same EER-threshold convention as Table~\ref{tab:evidence-card-triage}; the last entry is their union. The table does not evaluate an operational inspection policy.}
\label{tab:evidence-card-record-diagnostic}
\vspace{3pt}
\centering
\fontsize{6.3}{7.05}\selectfont
\setlength{\tabcolsep}{2.0pt}
\renewcommand{\arraystretch}{1.08}
\begin{tabularx}{0.94\linewidth}{@{}>{\raggedright\arraybackslash}p{0.18\linewidth}>{\raggedright\arraybackslash}p{0.22\linewidth}*{4}{>{\centering\arraybackslash}X}@{}}
\toprule
{\bfseries\boldmath Cue} & {\bfseries\boldmath Rule} & {\bfseries\boldmath Examples} & {\bfseries\boldmath \shortstack[c]{Record\\errors}} & {\bfseries\boldmath \shortstack[c]{Error cov.\\(\%)~$\uparrow$}} & {\bfseries\boldmath \shortstack[c]{Precision\\(\%)~$\uparrow$}} \\
\midrule
near record threshold & lowest 10\% $|s_{\mathrm{rec}}-\tau_{\mathrm{rec}}|$ & 408 & 165 & 47.97 & 40.44 \\
passive-record decision mismatch & passive and record decisions differ & 823 & 152 & 44.19 & 18.47 \\
retrieval-record decision mismatch & retrieval and record decisions differ & 564 & 131 & 38.08 & 23.23 \\
large record--retrieval gap & top 10\% $|f_{pw}-s_r|$ & 408 & 66 & 19.19 & 16.18 \\
any cue above & union of the four cues & 1,377 & 285 & 82.85 & 20.70 \\
\bottomrule
\end{tabularx}
\renewcommand{\arraystretch}{1.0}
\end{table}

The cue union is intended for triage, not for optimizing a single ranking rule.
Appendix Table~\ref{tab:record-review-load} compares it with scalar-only queues at the same review load.

\begin{table}[H]
\caption{Same-load comparison between scalar review queues and the four-cue union. Scalar queues rank examples by distance to their own EER threshold and are evaluated against errors from the calibrated record threshold rule on the matched set.}
\label{tab:record-review-load}
\vspace{3pt}
\centering
\fontsize{6.35}{7.10}\selectfont
\setlength{\tabcolsep}{2.1pt}
\renewcommand{\arraystretch}{1.08}
\begin{tabularx}{\linewidth}{@{}>{\raggedright\arraybackslash}p{0.18\linewidth}>{\raggedright\arraybackslash}X*{4}{>{\centering\arraybackslash}X}@{}}
\toprule
{\bfseries\boldmath Queue} & {\bfseries\boldmath Definition} & {\bfseries\boldmath Examples} & {\bfseries\boldmath \shortstack[c]{Share\\(\%)}} & {\bfseries\boldmath \shortstack[c]{Error cov.\\(\%)~$\uparrow$}} & {\bfseries\boldmath \shortstack[c]{Precision\\(\%)~$\uparrow$}} \\
\midrule
Passive margin & lowest $|s-\tau|$ at the cue-union review load & 1,377 & 33.75 & 32.85 & 8.21 \\
Retrieval margin & lowest $|s-\tau|$ at the cue-union review load & 1,377 & 33.75 & 40.99 & 10.24 \\
Fixed-rule margin & lowest $|s-\tau|$ at the cue-union review load & 1,377 & 33.75 & 77.03 & 19.24 \\
Learned-record margin & lowest $|s-\tau|$ at the cue-union review load & 1,377 & 33.75 & 88.66 & 22.15 \\
Four-cue union & union of the four cues & 1,377 & 33.75 & 82.85 & 20.70 \\
\bottomrule
\end{tabularx}
\renewcommand{\arraystretch}{1.0}
\end{table}

Appendix Table~\ref{tab:evidence-card-patterns} defines the selection criteria for the four worked records in Appendix Figure~\ref{fig:evidence-cards}.
The figure contrasts two correct decisions with a near-threshold error and a retrieval-driven error, making the retained fields visible without treating the examples as a deployment policy.

\begin{table}[H]
\caption{Selection criteria for the four decision-record examples in Figure~\ref{fig:evidence-cards}. Counts use the calibrated decision-record operating threshold on the 4,080 matched examples; C1--C2 are correct calibrated decisions and C3--C4 are calibrated-decision errors by construction.}
\label{tab:evidence-card-patterns}
\vspace{3pt}
\centering
\fontsize{6.3}{7.05}\selectfont
\setlength{\tabcolsep}{1.9pt}
\renewcommand{\arraystretch}{1.08}
\begin{tabularx}{0.96\linewidth}{@{}>{\centering\arraybackslash}p{0.07\linewidth}>{\raggedright\arraybackslash}p{0.28\linewidth}*{5}{>{\centering\arraybackslash}X}@{}}
\toprule
{\bfseries\boldmath Panel} & {\bfseries\boldmath Selection group} & {\bfseries\boldmath Examples} & {\bfseries\boldmath Share (\%)} & {\bfseries\boldmath \shortstack[c]{Group err.\\(\%)~$\downarrow$}} & {\bfseries\boldmath \shortstack[c]{Median\\margin}} & {\bfseries\boldmath $|f_{pw}-s_r|$} \\
\midrule
\textbf{C1} & \textbf{record--retrieval agreement} & \textbf{180} & \textbf{4.41} & \textbf{0.00} & \textbf{0.475} & \textbf{0.043} \\
\textbf{C2} & \textbf{retrieval-rescued spoof} & \textbf{146} & \textbf{3.58} & \textbf{0.00} & \textbf{0.242} & \textbf{0.609} \\
\textbf{C3} & \textbf{near-threshold error} & \textbf{77} & \textbf{1.89} & \textbf{100.00} & \textbf{0.042} & \textbf{0.463} \\
\textbf{C4} & \textbf{retrieval failure} & \textbf{41} & \textbf{1.00} & \textbf{100.00} & \textbf{0.305} & \textbf{0.600} \\
\bottomrule
\end{tabularx}
\renewcommand{\arraystretch}{1.0}
\end{table}

\begin{figure}[t]
\centering
\begin{tikzpicture}[
  font=\scriptsize\sffamily,
  casecard/.style={line width=0.68pt, rounded corners=5pt, minimum width=6.62cm, minimum height=3.50cm, inner sep=0pt, align=left},
  cardtitle/.style={font=\scriptsize\sffamily\bfseries, text=drawInk},
  statuslabel/.style={font=\scriptsize\sffamily\bfseries},
  cardtext/.style={font=\scriptsize\sffamily, text=drawInk},
  cardnote/.style={font=\scriptsize\sffamily\bfseries},
  barlabel/.style={font=\scriptsize\sffamily, text=drawInk},
  barvalue/.style={font=\scriptsize\sffamily, text=drawSlate},
  barframe/.style={draw=drawGrid!76, fill=white, line width=0.25pt}
]

\begin{scope}[shift={(-3.45,0.00)}]
\node[casecard, fill=drawTealFill, draw=drawTeal!82!black] (card) at (0,0) {};
\node[cardtitle, anchor=north west] at (-3.04,1.46) {\textbf{C1. fusion--retrieval agreement}};
\node[statuslabel, anchor=north east, text=drawTeal!82!black] at (3.04,1.46) {\textbf{correct}};
\node[cardtext, anchor=north west] at (-3.04,1.10) {truth: bona fide $\rightarrow$ bona fide};
\node[cardtext, anchor=north west] at (-3.04,0.82) {support: fold A15; mark present};
\node[cardtext, anchor=north west] at (-3.04,0.56) {nearest: NN bona fide ($d=31.1$)};

\node[barlabel, anchor=east] at (-2.70,0.10) {$s_p$};
\draw[barframe] (-2.62,0.06) rectangle ++(1.36,0.09);
\fill[drawNavy!82] (-2.62,0.06) rectangle ++(1.30,0.09);
\node[barvalue, anchor=west] at (-1.18,0.10) {0.96};

\node[barlabel, anchor=east] at (-2.70,-0.16) {$s_w$};
\draw[barframe] (-2.62,-0.21) rectangle ++(1.36,0.09);
\fill[drawAmber!88!black] (-2.62,-0.21) rectangle ++(0.41,0.09);
\node[barvalue, anchor=west] at (-1.18,-0.16) {0.30};

\node[barlabel, anchor=east] at (0.36,0.10) {$s_r$};
\draw[barframe] (0.44,0.06) rectangle ++(1.36,0.09);
\fill[drawTeal!88!black] (0.44,0.06) rectangle ++(0.95,0.09);
\node[barvalue, anchor=west] at (1.88,0.10) {0.70};

\node[barlabel, anchor=east] at (0.36,-0.16) {$s_m$};
\draw[barframe] (0.44,-0.21) rectangle ++(1.36,0.09);
\fill[drawPlum!78!black] (0.44,-0.21) rectangle ++(0.71,0.09);
\node[barvalue, anchor=west] at (1.88,-0.16) {0.53};
\node[cardtext, anchor=north west] at (-3.04,-0.58) {fixed 0.67; calibrated 1.00; bin 15/15};
\node[cardtext, anchor=north west] at (-3.04,-0.88) {gaps: $|s_p-s_w|$ 0.65; $|f_{pw}-s_r|$ 0.07};
\node[cardnote, anchor=north west, text=drawTeal!82!black] at (-3.04,-1.22) {cue: low record--retrieval gap};
\end{scope}

\begin{scope}[shift={(3.45,0.00)}]
\node[casecard, fill=drawTealFill, draw=drawTeal!82!black] (card) at (0,0) {};
\node[cardtitle, anchor=north west] at (-3.04,1.46) {\textbf{C2. retrieval-rescued spoof}};
\node[statuslabel, anchor=north east, text=drawTeal!82!black] at (3.04,1.46) {\textbf{correct}};
\node[cardtext, anchor=north west] at (-3.04,1.10) {truth: spoof $\rightarrow$ spoof};
\node[cardtext, anchor=north west] at (-3.04,0.82) {support: fold A12; mark present};
\node[cardtext, anchor=north west] at (-3.04,0.56) {nearest: NN A08 ($d=19.2$)};

\node[barlabel, anchor=east] at (-2.70,0.10) {$s_p$};
\draw[barframe] (-2.62,0.06) rectangle ++(1.36,0.09);
\fill[drawNavy!82] (-2.62,0.06) rectangle ++(1.36,0.09);
\node[barvalue, anchor=west] at (-1.18,0.10) {1.00};

\node[barlabel, anchor=east] at (-2.70,-0.16) {$s_w$};
\draw[barframe] (-2.62,-0.21) rectangle ++(1.36,0.09);
\fill[drawAmber!88!black] (-2.62,-0.21) rectangle ++(0.57,0.09);
\node[barvalue, anchor=west] at (-1.18,-0.16) {0.42};

\node[barlabel, anchor=east] at (0.36,0.10) {$s_r$};
\draw[barframe] (0.44,0.06) rectangle ++(1.36,0.09);
\fill[drawTeal!88!black] (0.44,0.06) rectangle ++(0.00,0.09);
\node[barvalue, anchor=west] at (1.88,0.10) {0.00};

\node[barlabel, anchor=east] at (0.36,-0.16) {$s_m$};
\draw[barframe] (0.44,-0.21) rectangle ++(1.36,0.09);
\fill[drawPlum!78!black] (0.44,-0.21) rectangle ++(0.68,0.09);
\node[barvalue, anchor=west] at (1.88,-0.16) {0.50};
\node[cardtext, anchor=north west] at (-3.04,-0.58) {fixed 0.35; calibrated 0.28; bin 5/15};
\node[cardtext, anchor=north west] at (-3.04,-0.88) {gaps: $|s_p-s_w|$ 0.58; $|f_{pw}-s_r|$ 0.71};
\node[cardnote, anchor=north west, text=drawTeal!82!black] at (-3.04,-1.22) {cue: retrieval rescue};
\end{scope}

\begin{scope}[shift={(-3.45,-3.98)}]
\node[casecard, fill=drawWarmFill, draw=drawAmber!88!black] (card) at (0,0) {};
\node[cardtitle, anchor=north west] at (-3.04,1.46) {\textbf{C3. near-threshold error}};
\node[statuslabel, anchor=north east, text=drawAmber!88!black] at (3.04,1.46) {\textbf{error}};
\node[cardtext, anchor=north west] at (-3.04,1.10) {truth: bona fide $\rightarrow$ spoof};
\node[cardtext, anchor=north west] at (-3.04,0.82) {support: fold A16; mark present};
\node[cardtext, anchor=north west] at (-3.04,0.56) {nearest: NN A07 ($d=24.6$)};

\node[barlabel, anchor=east] at (-2.70,0.10) {$s_p$};
\draw[barframe] (-2.62,0.06) rectangle ++(1.36,0.09);
\fill[drawNavy!82] (-2.62,0.06) rectangle ++(0.57,0.09);
\node[barvalue, anchor=west] at (-1.18,0.10) {0.42};

\node[barlabel, anchor=east] at (-2.70,-0.16) {$s_w$};
\draw[barframe] (-2.62,-0.21) rectangle ++(1.36,0.09);
\fill[drawAmber!88!black] (-2.62,-0.21) rectangle ++(0.39,0.09);
\node[barvalue, anchor=west] at (-1.18,-0.16) {0.29};

\node[barlabel, anchor=east] at (0.36,0.10) {$s_r$};
\draw[barframe] (0.44,0.06) rectangle ++(1.36,0.09);
\fill[drawTeal!88!black] (0.44,0.06) rectangle ++(0.41,0.09);
\node[barvalue, anchor=west] at (1.88,0.10) {0.30};

\node[barlabel, anchor=east] at (0.36,-0.16) {$s_m$};
\draw[barframe] (0.44,-0.21) rectangle ++(1.36,0.09);
\fill[drawPlum!78!black] (0.44,-0.21) rectangle ++(0.65,0.09);
\node[barvalue, anchor=west] at (1.88,-0.16) {0.48};
\node[cardtext, anchor=north west] at (-3.04,-0.58) {fixed 0.33; calibrated 0.49; bin 8/15};
\node[cardtext, anchor=north west] at (-3.04,-0.88) {gaps: $|s_p-s_w|$ 0.13; $|f_{pw}-s_r|$ 0.06};
\node[cardnote, anchor=north west, text=drawAmber!88!black] at (-3.04,-1.22) {cue: low margin};
\end{scope}

\begin{scope}[shift={(3.45,-3.98)}]
\node[casecard, fill=drawWarmFill, draw=drawAmber!88!black] (card) at (0,0) {};
\node[cardtitle, anchor=north west] at (-3.04,1.46) {\textbf{C4. retrieval failure}};
\node[statuslabel, anchor=north east, text=drawAmber!88!black] at (3.04,1.46) {\textbf{error}};
\node[cardtext, anchor=north west] at (-3.04,1.10) {truth: spoof $\rightarrow$ bona fide};
\node[cardtext, anchor=north west] at (-3.04,0.82) {support: fold A13; mark present};
\node[cardtext, anchor=north west] at (-3.04,0.56) {nearest: NN bona fide ($d=30.0$)};

\node[barlabel, anchor=east] at (-2.70,0.10) {$s_p$};
\draw[barframe] (-2.62,0.06) rectangle ++(1.36,0.09);
\fill[drawNavy!82] (-2.62,0.06) rectangle ++(0.09,0.09);
\node[barvalue, anchor=west] at (-1.18,0.10) {0.07};

\node[barlabel, anchor=east] at (-2.70,-0.16) {$s_w$};
\draw[barframe] (-2.62,-0.21) rectangle ++(1.36,0.09);
\fill[drawAmber!88!black] (-2.62,-0.21) rectangle ++(0.40,0.09);
\node[barvalue, anchor=west] at (-1.18,-0.16) {0.30};

\node[barlabel, anchor=east] at (0.36,0.10) {$s_r$};
\draw[barframe] (0.44,0.06) rectangle ++(1.36,0.09);
\fill[drawTeal!88!black] (0.44,0.06) rectangle ++(1.36,0.09);
\node[barvalue, anchor=west] at (1.88,0.10) {1.00};

\node[barlabel, anchor=east] at (0.36,-0.16) {$s_m$};
\draw[barframe] (0.44,-0.21) rectangle ++(1.36,0.09);
\fill[drawPlum!78!black] (0.44,-0.21) rectangle ++(0.68,0.09);
\node[barvalue, anchor=west] at (1.88,-0.16) {0.50};
\node[cardtext, anchor=north west] at (-3.04,-0.58) {fixed 0.59; calibrated 0.64; bin 10/15};
\node[cardtext, anchor=north west] at (-3.04,-0.88) {gaps: $|s_p-s_w|$ 0.23; $|f_{pw}-s_r|$ 0.82};
\node[cardnote, anchor=north west, text=drawAmber!88!black] at (-3.04,-1.22) {cue: large raw-score gap};
\end{scope}
\end{tikzpicture}%
\caption{Four retrospective decision-record examples. Each decision uses the calibrated operating threshold for the decision-record model; the fixed score is shown as a reference. Horizontal bars show passive (blue), probe (gold), retrieval (teal), and profile (purple) scores on the same 0--1 scale; each record also lists the decision status, watermark status, calibration bin, nearest-neighbor context, raw-score gaps, and diagnostic cue.}
\label{fig:evidence-cards}
\end{figure}
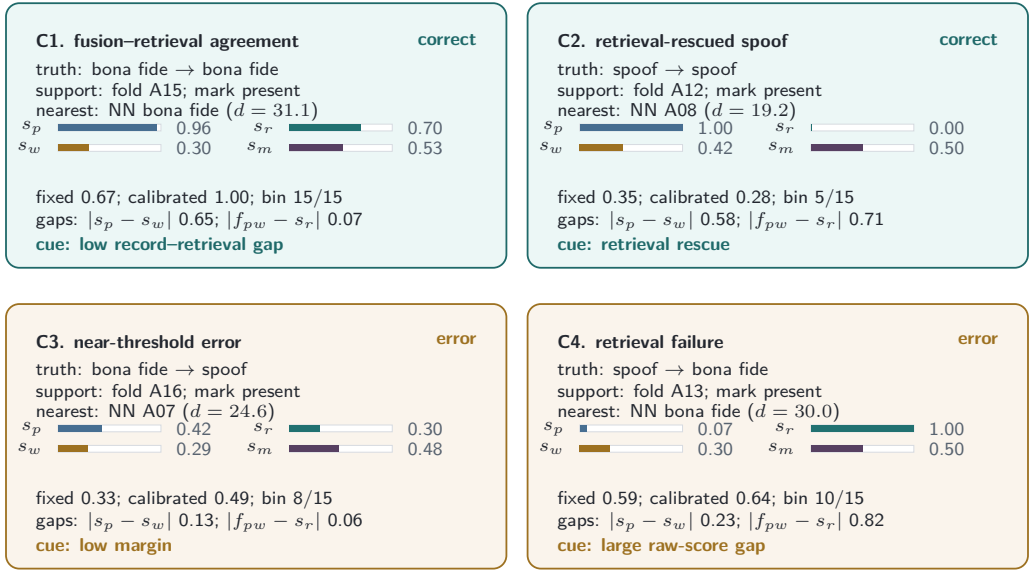

\begin{table}[H]
\caption{Paired bootstrap EER percentage-point deltas for Tables~\ref{tab:matched-results} and~\ref{tab:calibration-ablation}. Negative deltas favor the candidate. Intervals are nominal 95\% confidence intervals (CIs) and are not adjusted across comparisons.}
\label{tab:paired-deltas}
\centering
\fontsize{5.95}{6.70}\selectfont
\setlength{\tabcolsep}{0.8pt}
\renewcommand{\arraystretch}{1.06}
\begin{tabularx}{0.94\linewidth}{@{}>{\raggedright\arraybackslash}p{0.24\linewidth}@{\hspace{1pt}}>{\raggedright\arraybackslash}p{0.24\linewidth}@{\hspace{1pt}}>{\raggedright\arraybackslash}X@{\hspace{1pt}}>{\centering\arraybackslash}p{0.11\linewidth}@{}}
\toprule
{\bfseries\boldmath Baseline} & {\bfseries\boldmath Candidate} & {\bfseries\boldmath \shortstack[l]{Paired $\Delta$ EER (\%)\\[95\% CI]~$\downarrow$}} & {\bfseries\boldmath \shortstack[c]{Fold-level\\CI}} \\
\midrule
Passive CNN & \textbf{Fixed retrieval rule} & -9.24 [-11.89, -7.15] & -- \\
Passive + watermark & \textbf{Fixed retrieval rule} & -9.12 [-11.91, -7.06] & -- \\
Retrieval kNN & \textbf{Fixed retrieval rule} & -3.92 [-5.20, -3.22] & -- \\
Passive CNN & \textbf{Fixed retrieval + profile} & -8.19 [-10.43, -6.23] & -- \\
Passive + watermark & \textbf{Fixed retrieval + profile} & -8.07 [-10.52, -6.13] & -- \\
Retrieval kNN & \textbf{Fixed retrieval + profile} & -2.88 [-4.51, -1.51] & -- \\
Fixed retrieval-augmented rule & \textbf{Proposed decision record} & -3.48 [-4.61, -2.08] & -- \\
Cross-fit scalar fusion & \textbf{Proposed decision record} & -3.48 [-4.60, -2.34] & [-5.85, -0.20] \\
Retrieval + profile only & \textbf{Proposed decision record} & -8.21 [-9.83, -6.73] & [-12.23, -3.64] \\
Passive + retrieval only & \textbf{Proposed decision record} & -3.24 [-4.27, -2.08] & [-5.41, -0.24] \\
No-probe nonlinear control & \textbf{Proposed decision record} & -0.24 [-0.81, +0.46] & [-0.80, +0.49] \\
Passive margin only & \textbf{Proposed decision record} & -15.61 [-17.12, -14.11] & [-18.04, -13.95] \\
Passive shape control & \textbf{Proposed decision record} & -13.06 [-15.42, -10.89] & [-35.51, -2.49] \\
Passive-probe gap & \textbf{Proposed decision record} & -0.46 [-1.30, +0.14] & [-2.37, +0.11] \\
Shuffled-probe control & \textbf{Proposed decision record} & -0.34 [-1.26, +0.24] & [-2.18, +0.17] \\
Fusion-retrieval gap & \textbf{Proposed decision record} & -0.34 [-0.93, +0.22] & [-1.25, +0.25] \\
Squared-gap record & \textbf{Proposed decision record} & +0.10 [-0.26, +0.57] & [-0.26, +0.97] \\
Product-term control & \textbf{Proposed decision record} & -1.72 [-2.65, -0.67] & [-3.66, +0.12] \\
\bottomrule
\end{tabularx}
\renewcommand{\arraystretch}{1.0}
\end{table}

\begin{table}[H]
\caption{Paired expected calibration error (ECE) differences for named cross-fit controls on the 4,080 matched examples. Negative values favor the proposed decision record. Intervals are nominal 95\% intervals from 5,000 class-stratified bootstrap resamples and are not adjusted across contrasts.}
\label{tab:calibration-ece-intervals}
\vspace{3pt}
\centering
\fontsize{6.25}{7.05}\selectfont
\setlength{\tabcolsep}{2.0pt}
\renewcommand{\arraystretch}{1.06}
\begin{tabularx}{0.76\linewidth}{@{}>{\raggedright\arraybackslash}X>{\centering\arraybackslash}p{0.11\linewidth}>{\centering\arraybackslash}p{0.32\linewidth}@{}}
\toprule
{\bfseries\boldmath Control} & {\bfseries\boldmath ECE} & {\bfseries\boldmath Paired $\Delta$ ECE [95\% CI]~$\downarrow$} \\
\midrule
Cross-fit scalar fusion & 0.0840 & -0.0130 [-0.0159, -0.0090] \\
Passive + retrieval only & 0.0904 & -0.0195 [-0.0223, -0.0156] \\
No-probe nonlinear control & 0.0756 & -0.0047 [-0.0090, -0.0020] \\
Product-term control & 0.0780 & -0.0071 [-0.0093, -0.0035] \\
Squared-gap record & 0.0709 & +0.0001 [-0.0016, +0.0021] \\
\bottomrule
\end{tabularx}
\renewcommand{\arraystretch}{1.0}
\end{table}

\begin{table}[H]
\caption{Speaker-resampled EER deltas for key calibration controls. The candidate is the cross-fit absolute-gap model in Table~\ref{tab:calibration-ablation}. Negative deltas favor the candidate. Paired intervals resample matched examples, fold intervals resample the eight held-out calibration folds, and speaker intervals resample the 705 speaker IDs. Intervals are nominal 95\% and are not adjusted across comparisons.}
\label{tab:speaker-resampled-deltas}
\centering
\fontsize{6.8}{7.8}\selectfont
\setlength{\tabcolsep}{2.4pt}
\begin{tabular}{lrrr}
\toprule

{\bfseries\boldmath Baseline} & {\bfseries\boldmath Paired $\Delta$ EER} & {\bfseries\boldmath Fold CI} & {\bfseries\boldmath Speaker CI} \\
\midrule
Cross-fit scalar fusion & -3.48 [-4.60, -2.34] & [-5.85, -0.20] & [-4.63, -2.34] \\
No-probe nonlinear control & -0.24 [-0.81, +0.46] & [-0.80, +0.49] & [-0.82, +0.48] \\
Product-term control & -1.72 [-2.65, -0.67] & [-3.66, +0.12] & [-2.72, -0.65] \\
Squared-gap record & +0.10 [-0.26, +0.57] & [-0.26, +0.97] & [-0.28, +0.53] \\
\bottomrule
\end{tabular}
\end{table}

\begin{table}[H]
\caption{Per-fold EER for learned calibration on the 4,080 matched examples. Each held-out family is evaluated by a calibrator trained without spoof examples from that family. EER values are percentages; bold marks the best learned setting in that family.}
\label{tab:calibration-folds}
\centering
\fontsize{6.8}{7.8}\selectfont
\setlength{\tabcolsep}{3.0pt}
\begin{tabular}{lrrrrrr}
\toprule

{\bfseries\boldmath Family} & {\bfseries\boldmath Passive CNN} & {\bfseries\boldmath Fixed rule} & {\bfseries\boldmath Score fusion} & {\bfseries\boldmath Squared gaps} & {\bfseries\boldmath Nonlinear} & {\bfseries\boldmath Proposed model} \\
\midrule
A09 & 9.01 & 8.30 & 6.34 & 6.20 & \textbf{5.62} & 6.34 \\
A10 & 8.03 & 6.45 & 4.74 & 4.08 & \textbf{3.95} & 4.08 \\
A11 & 6.66 & 15.87 & 11.53 & \textbf{6.42} & 8.97 & \textbf{6.42} \\
A12 & 80.18 & 6.81 & 13.01 & \textbf{4.62} & 9.91 & 6.44 \\
A13 & 53.06 & 13.98 & 13.11 & \textbf{9.99} & 12.86 & \textbf{9.99} \\
A14 & 4.93 & 0.99 & \textbf{0.74} & 0.99 & \textbf{0.74} & 0.86 \\
A15 & 23.28 & 25.92 & 22.36 & \textbf{19.96} & \textbf{19.96} & 21.56 \\
A16 & 7.94 & 1.99 & \textbf{1.99} & \textbf{1.99} & \textbf{1.99} & \textbf{1.99} \\
\bottomrule
\end{tabular}
\end{table}

\begin{table}[H]
\caption{Calibration sensitivity to bona fide fold assignment. Spoof folds remain fixed by held-out synthesis family; only the stable hash that assigns bona fide examples to those folds changes. Each cell reports mean [minimum, maximum] over ten assignments; EER values are percentages.}
\label{tab:calibration-split-stability}
\vspace{3pt}
\centering
\fontsize{6.00}{6.80}\selectfont
\setlength{\tabcolsep}{1.1pt}
\renewcommand{\arraystretch}{1.05}
\begin{tabularx}{\linewidth}{@{}>{\raggedright\arraybackslash}p{0.24\linewidth}>{\centering\arraybackslash}p{0.08\linewidth}>{\centering\arraybackslash}X>{\centering\arraybackslash}X>{\centering\arraybackslash}X@{}}
\toprule
{\bfseries\boldmath Setting} & {\bfseries\boldmath Assign.} & {\bfseries\boldmath \mbox{Pooled EER (\%)~$\downarrow$}} & {\bfseries\boldmath \mbox{Fold EER (\%)~$\downarrow$}} & {\bfseries\boldmath ECE~$\downarrow$} \\
\midrule
Score fusion & 10 & 11.89\% [11.69, 12.05] & 9.45\% [9.27, 9.72] & \mbox{0.0840 [0.0836, 0.0845]} \\
Decision record (abs. gaps) & 10 & 8.42\% [8.33, 8.55] & 7.06\% [6.69, 7.31] & 0.0707 [0.0704, 0.0709] \\
Decision record (sq. gaps) & 10 & 8.22\% [8.11, 8.33] & 6.73\% [6.36, 7.14] & 0.0707 [0.0705, 0.0709] \\
\bottomrule
\end{tabularx}
\renewcommand{\arraystretch}{1.0}
\end{table}

The equal-width calibration view in Figure~\ref{fig:evidence-card-risk} shows that most miscalibration sits near the upper score range.
Appendix Table~\ref{tab:calibration-reliability} repeats the ECE comparison with equal-mass bins.

\begin{table}[H]
\caption{Calibration-bin sensitivity on the 4,080 matched examples. Width ECE is the main-table value; mass ECE uses 15 equal-mass bins of 272 examples. Fold-iso ECE applies an isotonic map fit on the other folds. Sparse bins count equal-width bins with fewer than 100 examples.}
\label{tab:calibration-reliability}
\centering
\fontsize{6.8}{7.8}\selectfont
\setlength{\tabcolsep}{1.2pt}
\begin{tabularx}{0.94\linewidth}{@{}>{\raggedright\arraybackslash}p{0.24\linewidth}*{6}{>{\centering\arraybackslash}X}@{}}
\toprule

{\bfseries\boldmath System} & {\bfseries\boldmath \shortstack[c]{Width\\ECE~$\downarrow$}} & {\bfseries\boldmath \shortstack[c]{Mass\\ECE~$\downarrow$}} & {\bfseries\boldmath \shortstack[c]{Fold-iso\\ECE~$\downarrow$}} & {\bfseries\boldmath \shortstack[c]{Sparse\\bins}} & {\bfseries\boldmath \shortstack[c]{Largest\\bin}} & {\bfseries\boldmath \shortstack[c]{Max mass\\gap}} \\
\midrule
Retrieval kNN & 0.1451 & 0.1451 & 0.0187 & 6 & 33.6\% & 0.4008 \\
Fixed retrieval-augmented rule & 0.2609 & 0.2557 & 0.0321 & 3 & 18.7\% & 0.4123 \\
\textbf{Proposed decision record} & \textbf{0.0709} & \textbf{0.0715} & \textbf{0.0178} & \textbf{6} & \textbf{49.3\%} & \textbf{0.2832} \\
\bottomrule
\end{tabularx}
\end{table}

\begin{table}[H]
\caption{Retrieval and profile sweep used to fix the matched comparison. Entries summarize retrieval and profile measurements before the stricter matched fusion join. The main table uses the HuBERT-large 20k default setting; WavLM settings provide sensitivity checks.}
\label{tab:retrieval-sweep}
\centering
\fontsize{6.8}{7.8}\selectfont
\setlength{\tabcolsep}{1.2pt}
\begin{tabularx}{0.98\linewidth}{@{}>{\raggedright\arraybackslash}p{0.30\linewidth}*{4}{>{\centering\arraybackslash}X}>{\raggedright\arraybackslash}X@{}}
\toprule

{\bfseries\boldmath Configuration} & {\bfseries\boldmath Support} & {\bfseries\boldmath Examples} & {\bfseries\boldmath \shortstack[c]{Retrieval\\EER (\%)}} & {\bfseries\boldmath \shortstack[c]{Profile\\EER (\%)}} & {\bfseries\boldmath \shortstack[c]{Worst family\\EER (\%)~$\downarrow$}} \\
\midrule
WavLM-large 20k default & 38,797 & 40,000 & 11.12 & 68.12 & A11 / 27.80 \\
WavLM-large 20k d6 & 38,797 & 40,000 & 11.30 & 38.77 & A11 / 21.55 \\
HuBERT-large 20k d6 & 38,797 & 40,000 & 16.27 & 43.29 & A15 / 26.13 \\
HuBERT-large 20k default (main) & 38,797 & 40,000 & 16.46 & 59.11 & A11 / 27.66 \\
WavLM-large 20k d2 & 38,797 & 40,000 & 17.23 & 68.83 & A11 / 27.08 \\
Wav2Vec2-large 20k d6 & 38,797 & 40,000 & 21.11 & 60.72 & A11 / 34.70 \\
Wav2Vec2-large 20k default & 38,797 & 40,000 & 21.50 & 63.32 & A11 / 34.52 \\
Wav2Vec2-large 20k d10 & 38,797 & 40,000 & 23.02 & 59.71 & A11 / 33.01 \\
HuBERT-base 40k default & 58,797 & 71,334 & 23.05 & 75.37 & A11 / 41.56 \\
WavLM-base+ 40k default & 58,797 & 71,334 & 24.57 & 71.93 & A11 / 48.88 \\
HuBERT-large 20k d8 & 38,797 & 40,000 & 26.04 & 43.72 & A15 / 35.18 \\
Wav2Vec2-base 40k default & 58,797 & 71,334 & 31.91 & 56.53 & A13 / 40.06 \\
WavLM-large 20k d10 & 38,797 & 40,000 & 43.79 & 53.30 & A13 / 53.87 \\
\bottomrule
\end{tabularx}
\end{table}

\begin{table}[H]
\caption{Stress checks for the auxiliary probe field and passive references on separate control sets. The detail column reports either WavLM expected calibration error (ECE) or the relevant probe-control summary. WM denotes watermark, Self-VC denotes self-voice conversion (VC), and CI denotes confidence interval.}
\label{tab:stress-diagnostics}
\vspace{3pt}
\centering
\fontsize{6.10}{6.95}\selectfont
\setlength{\tabcolsep}{1.2pt}
\begin{tabularx}{\linewidth}{@{}>{\raggedright\arraybackslash}p{0.14\linewidth}>{\raggedright\arraybackslash}p{0.16\linewidth}*{4}{>{\centering\arraybackslash}X}>{\raggedright\arraybackslash}p{0.16\linewidth}@{}}
\toprule
{\bfseries\boldmath Condition} & {\bfseries\boldmath Score} & {\bfseries\boldmath $n$} & {\bfseries\boldmath Balance} & {\bfseries\boldmath \mbox{EER (\%)~$\downarrow$}} & {\bfseries\boldmath \mbox{minDCF$_{\mathrm{BF}}$~$\downarrow$}} & {\bfseries\boldmath Detail} \\
\midrule
WavLM retrieval & same-run passive & 4,130 & 3,213/917 & 22.79 & 1.0000 & ECE 0.2868 \\
WavLM retrieval & same-run retrieval & 4,130 & 3,213/917 & 10.36 & 0.9947 & ECE 0.1867 \\
WavLM retrieval & same-run fixed rule & 4,130 & 3,213/917 & 8.27 & 0.9191 & ECE 0.2854 \\
Probe pair & WM probe & 20,000 & 10,000/10,000 & 0.00 & 0.0000 & paired originals \\
Source pair & WM probe & 2,000 & 1,000/1,000 & 50.00 & 0.9990 & CI [47.7,52.1]; 485 spk. \\
Self-VC & WM probe & 2,000 & 1,000/1,000 & 52.00 & 1.0000 & CI [49.6,54.1] \\
Self-VC & Passive CNN & 2,000 & 1,000/1,000 & 48.50 & 1.0000 & CI [46.3,50.8] \\
\bottomrule
\end{tabularx}
\end{table}

\begin{table}[H]
\caption{SSL attack stress across 24 attack variants per model. The noise\_snr0 variant uses a signal-to-noise ratio (SNR) of 0 dB. Mean EER averages the variants, and the final column names the worst variant observed for that model.}
\label{tab:attack-stress}
\centering
\vspace{3pt}
\fontsize{6.8}{7.8}\selectfont
\setlength{\tabcolsep}{2.4pt}
\begin{tabular}{@{}ll@{\hspace{12pt}}l@{\hspace{18pt}}l@{}}
\toprule
{\bfseries\boldmath Model} & {\bfseries\boldmath Runs} & {\bfseries\boldmath \mbox{Mean EER (\%)~$\downarrow$}} & {\bfseries\boldmath \mbox{Worst variant / EER (\%)~$\downarrow$}} \\
\midrule
HuBERT-large & 24 & 12.79 & noise\_snr0 / 31.22 \\
Wav2Vec2-large & 24 & 13.43 & noise\_snr0 / 35.96 \\
WavLM-large & 24 & 14.77 & highpass\_2500hz / 70.96 \\
\bottomrule
\end{tabular}
\end{table}

The perturbation ranking is consistent across the settings in Table~\ref{tab:attack-stress}; Figure~\ref{fig:attack-stress} shows the same ordering visually.

\begin{figure}[!t]
\centering
\makebox[\linewidth][c]{\resizebox{0.965\linewidth}{!}{%
\begin{tikzpicture}
\begin{axis}[
  drawAxis,
  width=0.93\linewidth,
  height=5.35cm,
  scale only axis,
  xmin=-0.08,
  xmax=3.08,
  ymin=18.0,
  ymax=27.2,
  xtick={0,1,2,3},
  xticklabels={100,500,1k,2k},
  ytick={18,20,22,24,26},
  xlabel={selected examples},
  ylabel={EER (\%, lower is better)},
  xlabel style={font=\small, yshift=0.2ex},
  ylabel style={font=\small, yshift=-0.4ex},
  tick label style={font=\small},
  ymajorgrids=true,
  xmajorgrids=true,
  grid style={draw=drawGrid!48,line width=0.28pt},
  axis line style={draw=drawSlate!66,line width=0.45pt},
  legend style={
    at={(0.5,1.04)},
    anchor=south,
    draw=drawGrid!70,
    fill=white,
    fill opacity=0.96,
    text opacity=1,
    rounded corners=1.2pt,
    font=\scriptsize,
    legend columns=3,
    /tikz/every even column/.append style={column sep=0.26cm},
  },
  legend image post style={line width=0.54pt,mark size=0.82pt},
  legend cell align={left},
  clip=false,
  every axis plot/.append style={line join=round, line cap=round},
  error bars/error bar style={line width=0.14pt,draw=drawSlate!52},
  error bars/error mark options={rotate=90,mark size=0.54pt,line width=0.14pt,draw=drawSlate!52},
]
\addplot+[line width=0.58pt, dash pattern=on 4.7pt off 2.0pt, mark=*, mark size=0.96pt, draw=plotBlue!84!black, mark options={fill=white, draw=plotBlue!84!black, line width=0.16pt}, error bars/.cd, y dir=both, y explicit] table[x expr=\coordindex,y=combined,y error=combined_std,col sep=comma] {figures/active_selection.dat};
\addlegendentry{combined}
\addplot+[line width=0.60pt, solid, mark=square*, mark size=0.98pt, draw=plotGreen!84!black, mark options={fill=plotGreen!62!white, draw=white, line width=0.16pt}, error bars/.cd, y dir=both, y explicit] table[x expr=\coordindex,y=disagreement,y error=disagreement_std,col sep=comma] {figures/active_selection.dat};
\addlegendentry{disagreement}
\addplot+[line width=0.58pt, dash pattern=on 5.8pt off 2.0pt, mark=triangle*, mark size=1.00pt, draw=plotOrange!86!black, mark options={fill=plotOrange!66!white, draw=white, line width=0.16pt}, error bars/.cd, y dir=both, y explicit] table[x expr=\coordindex,y=drift,y error=drift_std,col sep=comma] {figures/active_selection.dat};
\addlegendentry{drift}
\addplot+[line width=0.48pt, densely dotted, mark=diamond*, mark size=0.88pt, draw=plotGray!74, mark options={fill=white, draw=plotGray!74, line width=0.15pt}, error bars/.cd, y dir=both, y explicit] table[x expr=\coordindex,y=random,y error=random_std,col sep=comma] {figures/active_selection.dat};
\addlegendentry{random}
\addplot+[line width=0.56pt, dash pattern=on 2.8pt off 1.18pt on 0.75pt off 1.18pt, mark=pentagon*, mark size=0.94pt, draw=drawCoral!84!black, mark options={fill=white, draw=drawCoral!84!black, line width=0.16pt}, error bars/.cd, y dir=both, y explicit] table[x expr=\coordindex,y=uncertainty,y error=uncertainty_std,col sep=comma] {figures/active_selection.dat};
\addlegendentry{uncertainty}
\node[anchor=west,font=\scriptsize,text=drawSlate!90] at (axis cs:0.05,18.55) {lower is better};
\end{axis}
\end{tikzpicture}%
}}
\caption{Active-label selection over repeated adaptation seeds. Curves show mean leave-family-out EER after adapting with each label budget; vertical bars show one standard deviation across seeds. The curves cross, so no acquisition rule is uniformly best across budgets.}
\label{fig:active-selection}
\end{figure}
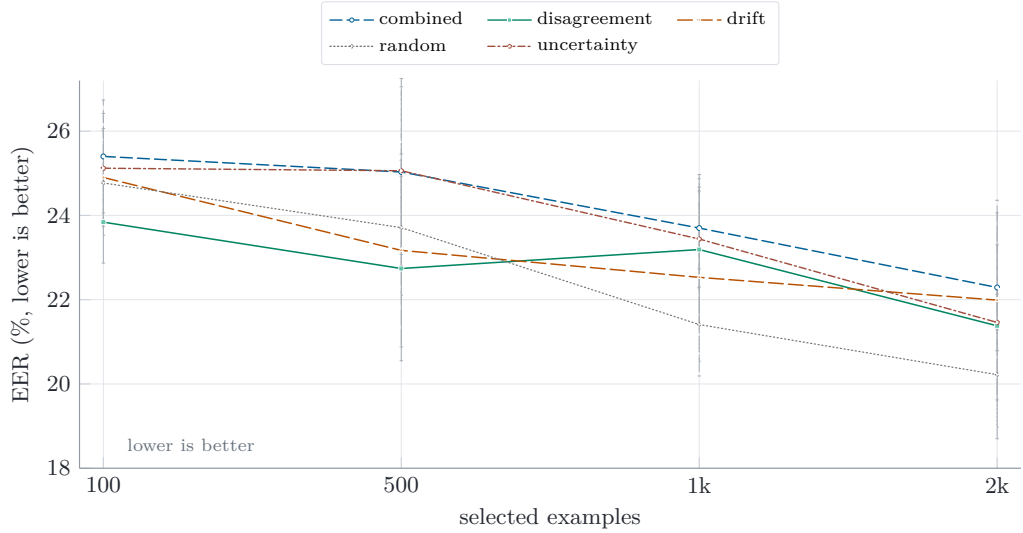

Across six repeat seeds, Figure~\ref{fig:active-selection} repeats the adaptation experiment after queried ASVspoof~5 development examples are appended to the training split; evaluation uses the remaining development examples.
Disagreement is best at the two smallest budgets, whereas random selection is best at the two largest budgets.
At 100 and 500 queried labels, disagreement lowers EER relative to random by 0.93 and 0.97 percentage points, respectively.
The ordering reverses at larger budgets in Figure~\ref{fig:active-selection}.
Because the ordering depends on budget, we treat active selection as a separate experimental-design question, not as an extension of the matched-subset result.

\input{figures/attack_stress.tex}

Figure~\ref{fig:attack-stress} shows that frozen SSL detectors remain near the clean benchmark under low-frequency and mild-resampling changes, but fail most under noise, quantization, and 2.5 kHz high-pass filtering.
Table~\ref{tab:attack-stress} reports mean and worst-case EER outside the main matched-subset result.

\end{document}